\documentstyle[preprint,aps,epsfig]{revtex}
\newcommand{\bq}{\begin{quotation}}
\newcommand{\eq}{\end{quotation}}
\newcommand{\be}{\begin{equation}}
\newcommand{\ee}{\end{equation}}

\begin{document}
\title{Early History of Gauge Theories and Kaluza-Klein\\
Theories, with a Glance at Recent Developments}
\author{Lochlain O'Raifeartaigh}
\address{Dublin Institute for Advanced Studies, \\
Dublin 4, Ireland}
\author{Norbert Straumann}
\address{Institut f\"ur Theoretische Physik der
Universit\"at Z\"urich--Irchel,\\
Z\"urich,
Switzerland}
\maketitle
\begin{abstract}One of the major developments of twentieth century
physics has been the gradual recognition that a common feature of the known
fundamental interactions is their gauge structure. In this article
the authors review the early history of gauge theory, from Einstein's theory
of gravitation to the appearance of non-abelian gauge theories in the
fifties. The authors also review the early history of dimensional reduction,
which played an important role in the developement of gauge-theory.
A description is given how, in recent times, the ideas of gauge-theory
and dimensional reduction have emerged naturally in the context of string
theory and non-commutative geometry.
\end{abstract}
\tableofcontents

\section{Introduction}

It took decades until physicists understood that all known
fundamental interactions can be described in terms of gauge theories.
Our historical account begins with Einstein's general theory of
relativity (GR), which is a non-Abelian gauge theory of a special 
type (see Secs. 3,7). The other gauge theories emerged in a slow
 and complicated process gradually from GR, and their common geometrical
structure --- best expressed in terms of connections of fiber bundles
--- is now widely recognized. Thus, 
H. Weyl was right when he wrote in the preface to the first edition 
of Space -- Time -- Matter (RZM) early in 1918: 
``Wider expanses and greater depths are now exposed to the 
searching eye of knowledge, regions of  which  we  had  not  even  a 
presentiment. 
It has brought us much nearer to grasping the plan that
underlies all physical happening'' \cite{1}. 

It was Weyl himself who made in 1918 the first attempt to extend GR in order 
to describe gravitation and electromagnetism within a unifying
geometrical framework \cite{2}. This brilliant proposal contains 
the germs of all mathematical aspects of a non-Abelian gauge theory,
as we will make clear in Sec. 2. The words gauge (Eich--) transformation
and gauge invariance appear for the first time in this paper, but in the 
everyday meaning of change of length or change of calibration\footnote{
The German word ``eichen'' probably comes from the Latin ``aequare'', i.e.,
equalizing the length to a standard one.}.

Einstein admired Weyl's theory as  ``a coup of genius of the first rate 
\ldots '',
but immediately realized that it was physically untenable: ``Although your 
idea is so 
beautiful, I have to declare frankly that, in my opinion, it is impossible 
that 
the theory corresponds to nature.'' This led to an intense exchange of 
letters
between Einstein (in Berlin) and Weyl (at the ETH in Z\"urich), part of which 
has now been published in vol.~8 of {\sl The Collected Papers} of Einstein \cite{vol8}.
(The article \cite{3} gives an account of this correspondence which 
is preserved in the Archives of the ETH.) No agreement was reached, but
Einstein's intuition proved to be right.

Although Weyl's attempt was a failure as a physical theory it paved the way
for the correct understanding of gauge invariance. Weyl himself re-interpreted
his original theory after the advent of quantum theory in a seminal paper 
\cite{4}
which we will discuss at length in Sec. 3. Parallel developments by other
workers
and interconnections are indicated in Fig.1. 

At the time Weyl's contributions to theoretical physics were not appreciated 
very 
much, since they did not really add new physics. The attitude of the leading
theoreticians is expressed with familiar bluntness in a letter by Pauli to 
Weyl
of July 1, 1929, after he had seen a preliminary account of Weyl's work:
\begin{quotation}

{\sl Before me lies the April edition of the Proc.Nat.Acad. (US). 
Not only does it contain an article from you under ``Physics''
but shows that you are now in a `Physical Laboratory': from what I hear 
you have even been given a chair in `Physics' in America. 
I admire your courage; since the conclusion is inevitable that you 
wish to be judged, not for success in pure mathematics, but for your
true but unhappy love for physics \cite{5}.}
\end{quotation}

Weyl's reinterpretation of his earlier speculative proposal had actually 
been suggested 
before by London and Fock, but it was Weyl who emphasized the role of gauge 
invariance as a {\em symmetry principle} from which electromagnetism 
can be {\em derived}. It took several decades until the importance 
of this symmetry principle --- in its generalized form to non-Abelian
gauge groups developed by Yang, Mills, and others --- became also 
fruitful for a description of the weak and strong interactions. 
The mathematics of the non-Abelian generalization of Weyl's 1929 paper
would have been an easy task for a mathematician of his rank, but at the time
there was no motivation for this from the physics side. The known properties
of the weak and strong nuclear interactions, in particular their short range,
did not point to a gauge theoretical description. We all know that the gauge
symmetries of the Standard Model are very hidden and it is, therefore,
not astonishing that progress was very slow indeed.

In this paper we present only the history up to the invention of
Yang-Mills theory in 1954. The independent discovery of this theory
by other authors has already been described in Ref.~\cite{6}. Later
history covering the application of the Yang-Mills theory to the electroweak
and strong interactions is beyond our scope. The main features of these
applications are well-known and are covered in contemporary text-books.
A modern aspect that we do wish to mention, however, is the emergence
of both gauge-theory and dimensional reduction in two fields other than
traditional quantum field theory, namely string theory and non-commutative
geometry, as their emergence in these fields is a natural extension of the
early history.
Indeed in string theory both gauge-invariance and dimensional reduction
occur in such a natural way that it is probably not an exaggeration to
say that, had they not been found earlier, they would have been discovered
in this context. The case of non-commutative geometry is a little different
as the gauge-principle is used as an input, but the change from a continuum
to a discrete structure produces qualitatively new features. Amongst
these is an interpretation of the Higgs field as a gauge-potential and
the emergence of a dimensional reduction that avoids the usual embarrassment
concerning the fate of the extra dimensions.

\begin{figure}\label{fig1}
\begin{center}
\epsfig{file=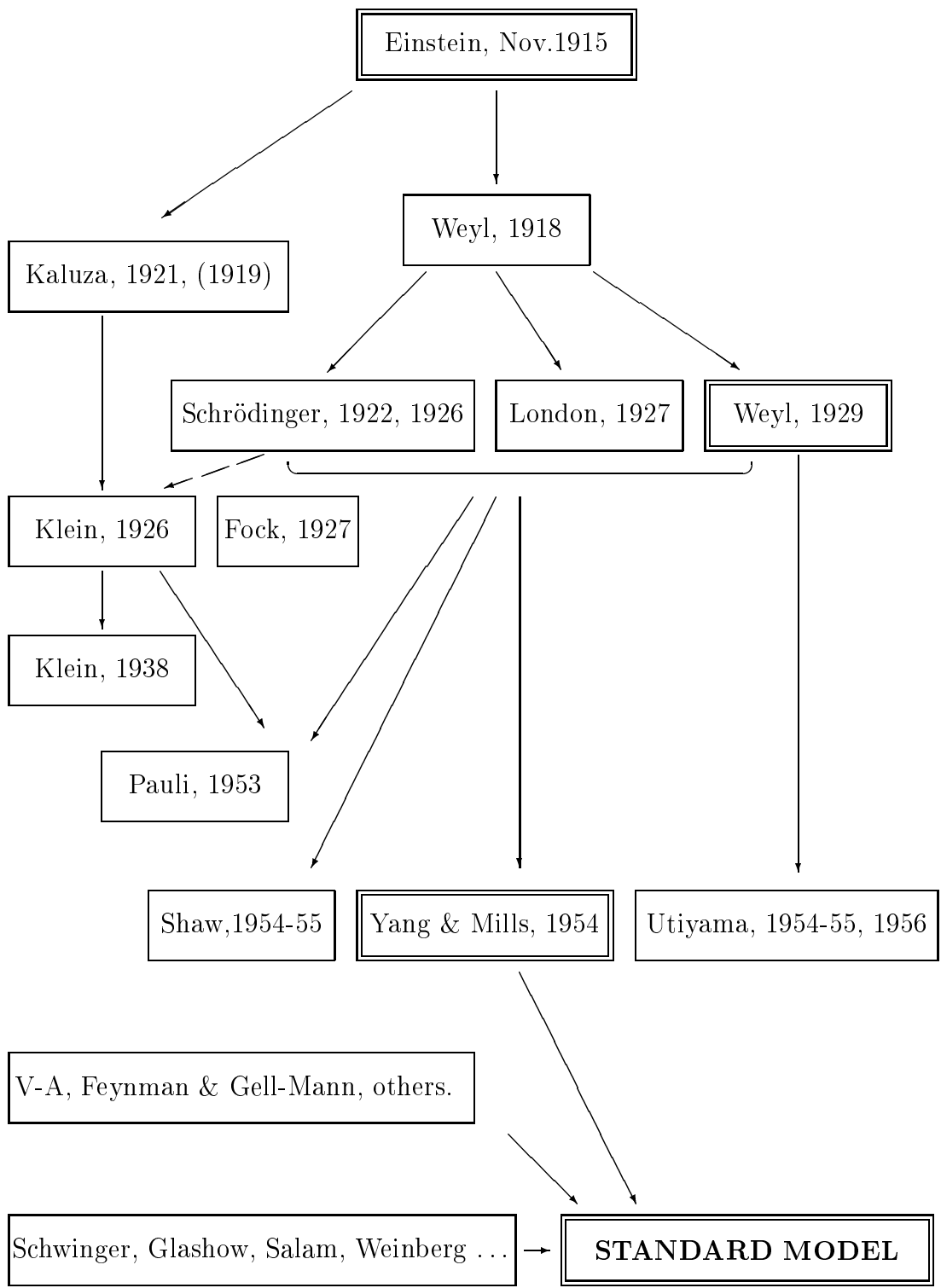,height=20cm}
\caption{Key papers in the development of gauge theories.}
\end{center}
\end{figure}

This historical account is not intended to be comprehensive, but should be
regarded as a supplement to the book~\cite{6}, published by one of us. There
one finds also English translations of the most important papers of the
early period, as well as Pauli's letters to Pais on non-Abelian Kaluza-Klein
reductions. These works underlie the diagram in Fig.~\protect\ref{fig1}.


\clearpage

\section{Weyl's Attempt to Unify Gravitation
\protect\newline
and Electromagnetism}

On the 1$^{st}$ of March 1918 Weyl writes in a letter to Einstein:
``These days I succeeded, as I believe, to derive electricity and gravitation
from a common source \ldots ''. Einstein's prompt reaction by postcard
indicates already a physical objection which he explained in detail shortly
afterwards. Before we come to this we have to describe Weyl's theory of 1918.

\subsection{Weyl's Generalization of Riemannian Geometry}

Weyl's starting point was purely mathematical. He felt a certain uneasiness
about Riemannian geometry, as is clearly expressed by the following sentences 
early
in his paper:
\begin{quotation}
{\sl But in Riemannian geometry described above there is contained a last
element of geometry ``at a distance''  (ferngeometrisches Element) --- with no good
reason, as far as I can see; it is due only to the accidental development 
of Riemannian geometry from Euclidean geometry. The metric allows the
two magnitudes of two vectors to be compared, not only at the same point, but 
at any
arbitrarily separated points.} {\it A true infinitesimal geometry should, 
however,
recognize only a principle for transferring the magnitude of a vector to an
 infinitesimally close point} {\sl and then, on transfer to an arbitrary 
distant point, the integrability of the magnitude of a vector is no more to be 
expected that the
integrability of its direction.}
\end{quotation}

After these remarks Weyl turns to physical speculation and continues as follows:

\begin{quotation}
{\sl On the removal of this inconsistency there appears a geometry that,
surprisingly,
when applied to the world,} {\it explains not only the gravitational
phenomena but 
also the electrical.} {\sl According to the resultant theory 
both spring from the same source, indeed} {\it in general one 
cannot separate gravitation and electromagnetism in a unique manner}.
{\sl In this theory} {\it all physical
quantities have a world geometrical meaning; the action appears from the 
beginning as a pure number. It leads to an essentially unique universal law;
it even allows us to understand in a certain sense why the world is
four-dimensional}.
\end{quotation}

In brief, Weyl's geometry can be described as follows
(see also ref.~\cite{13}). First, the spacetime 
manifold $M$ is equipped with a conformal structure, i.e.,
with a class $[g]$ of
conformally equivalent Lorentz metrics $g$ (and not a
definite metric as in GR).
This corresponds to the requirement that it should only
be possible to compare
lengths at one and the same world point. Second, it is
assumed, as in 
Riemannian geometry, that there is an affine 
(linear) torsion-free connection 
which defines a covariant derivative $\nabla$, and
respects the conformal structure.
Differentially this means that for any $g\in[g]$ the
covariant derivative $\nabla g$
should be proportional to $g$:
\be                                                \label{2.1}
\nabla g =-2A\otimes g\ \ \ \ \ \ \ 
(\nabla_{\lambda}g_{\mu\nu}=-2A_{\lambda}g_{\mu\nu}),
\ee
where $A=A_{\mu}dx^{\mu}$ is a differential 1-form.

Consider now a curve $\gamma: [0,1]\rightarrow M$ and a
parallel-transported
vector field $X$ along $\gamma$. If $l$ is the length of $X$,
measured with a representative $g\in[g]$, we obtain from (\ref{2.1})
the following relation between $l(p)$ for the initial point
$p=\gamma(0)$ and $l(q)$ for the end point $q=\gamma(1)$:
\be                                                \label{2.2}
l(q)=\exp\left(-\int_{\gamma}A\right)\ l(p).
\ee
Thus, the ratio of lengths in $q$ and $p$ (measured with
$g\in[g]$) {\it depends in general on the connecting path $\gamma$}
(see Fig.2). The length is only independent of $\gamma$ if the
curl of $A$,
\be                                                \label{2.3}
F=dA\ \ \ \  \ \ \ \ (F_{\mu\nu}=\partial_{\mu}A_{\nu}-
\partial_{\nu}A_{\mu}),
\ee
vanishes.
\begin{figure}\label{fig2}
\begin{center}
\epsfig{file=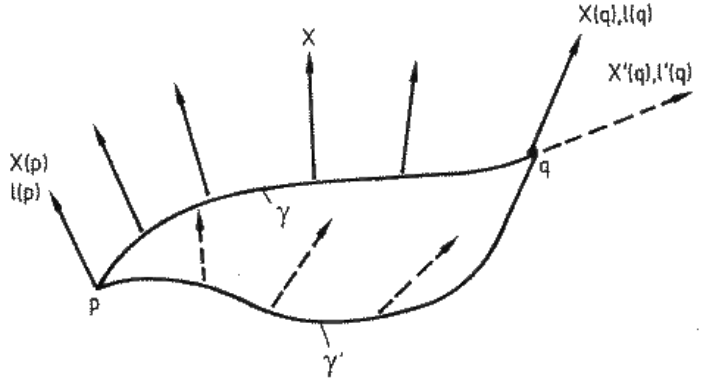,width=10cm}
\caption{Path dependence of parallel displacement and transport 
of length in  Weyl space.}
\end{center}
\end{figure}

The compatibility requirement (\ref{2.1}) leads to the following
expression for the Christoffel symbols in Weyl's geometry:
\be                                                \label{2.4}
\Gamma^{\mu}_{\nu\lambda}=\frac{1}{2}g^{\mu\sigma}(
g_{\lambda\sigma,\nu}+g_{\sigma\nu,\lambda}-g_{\nu\lambda,\sigma})
+g^{\mu\sigma}(g_{\lambda\sigma}A_{\nu}+g_{\sigma\nu}A_{\lambda}-
g_{\nu\lambda}A_{\sigma}).
\ee
The second $A$-dependent term is a characteristic new piece
in Weyl's geometry which has to be added to the Christoffel symbols
of Riemannian geometry.

Until now we have chosen a fixed, but arbitrary metric in the
conformal class $[g]$. This corresponds to a choice of calibration
(or gauge). Passing to another calibration with metric $\bar{g}$,
related to $g$ by
\be                                                \label{2.5}
\bar{g}=e^{2\lambda}g,
\ee
the potential $A$ in (\ref{2.1}) will also change to $\bar{A}$, say.
Since the covariant derivative has an absolute meaning,
$\bar{A}$ can easily be worked out: On the one hand we have by
definition
\be                                                
\nabla \bar{g} =-2\bar{A}\otimes\bar{g}, 
\ee
and on the other hand we find for the left side with (\ref{2.1})
\be                                                
\nabla\bar{g}=\nabla(e^{2\lambda}g)=
2d\lambda\otimes\bar{g}+e^{2\lambda}\nabla g=
2d\lambda\otimes\bar{g}-2A\otimes\bar{g}.
\ee
Thus
\be                                                \label{2.6}
\bar{A}=A- d\lambda\ \ \ \ \ \
(\bar{A}_{\mu}=A_{\mu}-\partial_{\mu}\lambda).
\ee
This shows that a change of calibration of the metric induces a
{\it ``gauge transformation''} for $A$:
\be                                                \label{2.7}
g\rightarrow e^{2\lambda}g,\ \ \ \
A\rightarrow A-d\lambda.
\ee
Only gauge classes have an absolute meaning.
(The Weyl connection is, however, gauge-invariant. This is conceptually
clear, but can also be verified by direct calculation from
expression~Eq.(\ref{2.4}).)

\subsection{Electromagnetism and Gravitation}

Turning to physics, Weyl assumes that his ``purely
infinitesimal geometry'' describes the structure of spacetime
and consequently he requires that physical laws should satisfy
a double-invariance: 1. They must be invariant with respect to
arbitrary smooth coordinate transformations.
2. They must be {\it gauge invariant}, i.e.,
invariant with respect to substitutions (\ref{2.7})
for an arbitrary smooth function $\lambda$.

Nothing is more natural to Weyl, than identifying $A_{\mu}$
with the vector potential and $F_{\mu\nu}$ in eq.(\ref{2.3})
with the field strength of electromagnetism.
In the absence of electromagnetic fields ($F_{\mu\nu}=0$)
the scale factor $\exp(-\int_{\gamma}A)$ in (\ref{2.2})
for length transport becomes path independent (integrable)
and one can find a gauge such that $A_{\mu}$ vanishes for
simply connected spacetime regions.
In this special case one is in the same situation as in GR.

Weyl proceeds to find an action which is generally invariant
as well as gauge invariant and which would give the coupled field
equations for $g$ and $A$. We do not want to enter into this,
except for the following remark. In his first paper \cite{2}
Weyl proposes what we call nowadays the Yang-Mills action
\be                                                   \label{2.8}
S(g,A)=-\frac{1}{4}\int Tr(\Omega\wedge\ast\Omega).
\ee
Here $\Omega$ denotes the curvature form and $\ast\Omega$
its Hodge dual\footnote{The integrand in (\ref{2.8}) is in
local coordinates indeed just the expression
$R_{\alpha\beta\gamma\delta} R^{\alpha\beta\gamma\delta}
\sqrt{-g}dx^{0}\wedge\ldots\wedge dx^{3}$ which is used
by Weyl ($R_{\alpha\beta\gamma\delta}$ $=$ curvature tensor of the
Weyl connection).}.
Note that the latter is gauge invariant, i.e., independent of the
choice of $g\in[g]$. In Weyl's geometry the curvature form
splits as $\Omega=\hat{\Omega}+F$, where $\hat{\Omega}$ is the
metric piece \cite{13}. Correspondingly, the action also splits,
\be                                                  \label{2.9}
Tr (\Omega\wedge\ast\Omega) =
Tr (\hat{\Omega}\wedge\ast\hat{\Omega})
+F\wedge\ast F.
\ee
The second term is just the Maxwell action. Weyl's theory
thus contains formally all aspects of a non-Abelian gauge theory.

Weyl emphasizes, of course, that the Einstein-Hilbert
action is not gauge invariant. Later work by Pauli \cite{14}
and by Weyl himself \cite{1,2} led soon to the conclusion that
the action (\ref{2.8}) could not be the correct one, and other
possibilities were investigated (see the later editions of RZM).

Independent of the precise form of the action Weyl shows that in
his theory gauge invariance implies the {\it conservation of electric
charge} in much the same way as general coordinate invariance
leads to the conservation of energy and momentum\footnote{We adopt here the somewhat naive interpretation of
energy-momentum conservation for generally invariant theories
of the older literature.}.
This beautiful connection pleased him particularly:
``\ldots [it] seems to me to be the strongest general argument
in favour of the present theory --- insofar as it is permissible
to talk of justification in the context of pure speculation.''
The invariance principles imply five `Bianchi type' identities.
Correspondingly, the five conservation laws follow in two
independent ways from the coupled field equations
and may be ``termed the eliminants'' of the latter. These structural connections
hold also in modern gauge theories.

\subsection{Einstein's Objection and Reactions of Other Physicists}

After this sketch of Weyl's theory we come to Einstein's
striking counterargument which he first communicated to Weyl
by postcard (see Fig.~3). The problem is that if the idea of a
nonintegrable length connection (scale factor) is correct, 
then the behavior of clocks would depend on their history.
Consider two identical atomic clocks in adjacent world points
and bring them along different world trajectories which
meet again in adjacent world points.
According to (\ref{2.2}) their frequencies would
then generally differ. This is in clear contradiction with
empirical evidence, in particular with the existence of stable
atomic spectra. Einstein therefore concludes (see \cite{3}):
\begin{quotation}
{\sl \ldots (if) one drops the connection of the $ds$ to the
measurement of distance and time, then relativity looses all its
empirical basis.}
\end{quotation}

Nernst shared Einstein's objection and demanded on behalf of the
Berlin Academy that it should be printed in a short amendment to Weyl's
article. Weyl had to accept this. One of us has described the
intense and instructive subsequent correspondence between Weyl
and Einstein elsewhere \cite{3} (see also Vol.~8 of Ref.~\cite{vol8}). As an example, let us quote from
one of the last letters of Weyl to Einstein:
\begin{quotation}
{\sl This [insistence] irritates me of course, because
experience has proven that one can rely on your intuition;
so unconvincing as your counterarguments seem to me, as
I have to admit \ldots}
\end{quotation}

\begin{quotation}
{\sl By the way, you should not believe that I was driven to
introduce the linear differential form in addition to the
quadratic one by physical reasons. I wanted, just to the
contrary, to get rid of this `methodological inconsistency
{\it (Inkonsequenz)}' which has been a bone of contention
to me already much earlier. And then, to my surprise, I realized
that it looked as if it might explain electricity. You clap
your hands above your head and shout: But physics is not
made this way ! (Weyl to Einstein 10.12.1918).}
\end{quotation}

Weyl's reply to Einstein's criticism was, generally speaking, this:
The real behavior of measuring rods and clocks (atoms and atomic systems)
in arbitrary electromagnetic and gravitational fields can be
 deduced only from 
a dynamical theory of matter. 

Not all leading physicists reacted negatively. Einstein
transmitted a very positive first reaction by Planck, and
Sommerfeld wrote enthusiastically to Weyl that there was
``\ldots hardly doubt, that you are on the correct path and not
on the wrong one.''

In his encyclopedia article on relativity \cite{15} Pauli gave
a lucid and precise presentation of Weyl's theory, but
commented on Weyl's point of view very critically. At the end he
states:
\begin{quotation}
{\sl \ldots In summary one may say that Weyl's theory has not
yet contributed to getting closer to the solution of the problem
of matter.}
\end{quotation}

Also Eddington's reaction was at first very positive but he
soon changed his mind and denied the physical relevance of
Weyl's geometry.

The situation was later appropriately summarized by F.London
in his 1927 paper \cite{16} as follows:
\begin{quotation}
{\sl In the face of such elementary experimental evidence,
it must have been an unusually strong metaphysical conviction
that prevented Weyl from abandoning the idea that Nature would
have to make use of the beautiful geometrical possibility
that was offered. He stuck to his conviction and evaded
discussion of the above-mentioned contradictions through a rather
unclear re-interpretation of the concept of ``real state'',
which, however, robbed his theory of its immediate physical
meaning and attraction.}
\end{quotation}
In this remarkable paper, London suggested a reinterpretation of Weyl's
principle of gauge invariance within the new quantum mechanics: The role
of the metric is taken over by the wave function, and the rescaling of the
metric has to be replaced by a phase change of the wave function.

In this context an astonishing early paper by Schr\"odinger~\cite{11n}
has to be mentioned, which also used Weyl's ``World Geometry'' and is
related to Schr\"odinger's later invention of wave mechanics. This relation
was discovered by V.~Raman and P.~Forman~\cite{12n}. (See also the discussion
by C.N.~Yang in~\cite{13n}.)

Simultaneously with London, V.~Fock~\cite{14n} arrived along a completely
different line at the principle of gauge invariance in the framework
of wave mechanics. His approach was similar to the one by O.~Klein, which
will later be discussed in detail (Sect.~IV).

The contributions by Schr\"odinger~\cite{11n}, London~\cite{16} and
Fock~\cite{14n}
are commented in~\cite{6}, where also English translations of the original
papers can be found. Here, we concentrate on Weyl's seminal paper
``Electron and Gravitation''.
\begin{figure}\label{fig3}
\epsfig{file=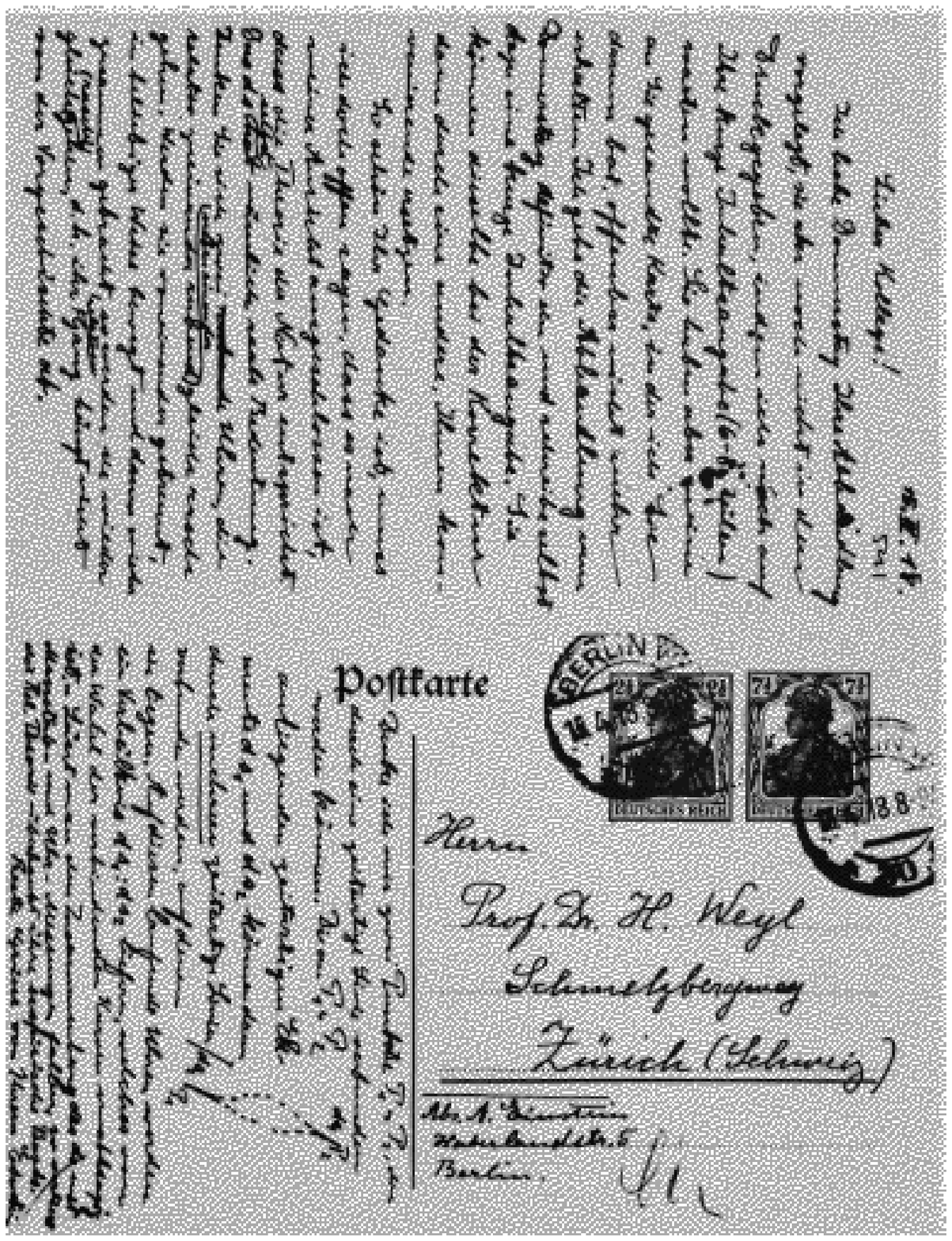,width=16cm}
\caption{Postcard of Einstein to Weyl 15.4.1918 (Archives of ETH).}
\end{figure}
\clearpage

\section{Weyl's 1929 Classic: ``Electron and Gravitation''}

Shortly before his death late in 1955, Weyl wrote for his
{\it Selecta} \cite{17} a postscript to his early attempt in
1918 to construct a `unified field theory'. There he expressed
his deep attachment to the gauge idea and adds (p.192):
\begin{quotation}
{\sl  Later the quantum-theory introduced the Schr\"odinger-Dirac
potential $\psi$ of the electron-positron field; it carried with it
an experimentally-based principle of gauge-invariance which
guaranteed the conservation of charge, and connected the $\psi$
with the electromagnetic potentials $\phi_{i}$ in the same way
that my speculative theory had connected the gravitational
potentials $g_{ik}$ with the $\phi_{i}$, and measured the
$\phi_{i}$ in known atomic, rather than unknown cosmological units.
I have no doubt but that the correct context for the principle
of gauge-invariance is here and not, as I believed in 1918, in the
intertwining of electromagnetism and gravity.}
\end{quotation}

This re-interpretation was developed by Weyl in one of the great
papers of this century \cite{4}. Weyl's classic does not only
give a very clear formulation of the gauge principle, but contains,
in addition, several other important concepts and results ---
in particular his two-component spinor theory. The richness and scope 
of the paper is
clearly visible from the following table of contents:
\begin{quotation}
{\sl Introduction. Relationship of General Relativity to the
quantum-theoretical field equations of the spinning electron:
mass, gauge-invariance, distant-parallelism. Expected modifications
of the Dirac theory. -I. Two-component theory: the wave function
$\psi$ has only two components. -$\S 1$. Connection between the
transformation of the $\psi$ and the transformation of a normal
tetrad  in four-dimensional space. Asymmetry of past and future,
of left and right. -$\S 2$. In General Relativity the metric
at a given point is determined by a normal tetrad. Components
of vectors relative to the tetrad and coordinates. Covariant
differentiation of $\psi$. -$\S 3$. Generally invariant form
of the Dirac action, characteristic for the wave-field of matter.
-$\S 4$. The differential conservation law of energy and momentum
and the symmetry of the energy-momentum tensor as a consequence
of the double-invariance (1) with respect to coordinate
transformations (2) with respect to rotation of the tetrad.
Momentum and moment of momentum for matter. -$\S 5$. Einstein's
classical theory of gravitation in the new analytic formulation.
Gravitational energy. -$\S 6$. The electromagnetic field. From
the arbitrariness of the gauge-factor in $\psi$ appears the necessity
of introducing the electromagnetic potential. Gauge invariance
and charge conservation. The space-integral of charge. The
introduction of mass. Discussion and rejection of another possibility
in which electromagnetism appears, not as an accompanying
phenomenon of matter, but of gravitation.}
\end{quotation}

The modern version of the gauge principle is already spelled out
in the introduction:

\begin{quotation}
{\sl The Dirac field-equations for $\psi$ together with the Maxwell
equations for the four potentials $f_{p}$ of the electromagnetic field
have an invariance property which is formally similar to the one which
I called gauge-invariance in my 1918 theory of gravitation and
electromagnetism; the equations remain invariant when one makes
the simultaneous substitutions
$$\psi\ \ \ {\rm by}\ \ \ e^{i\lambda}\psi\ \ \ \ {\rm and}\ \ \
f_{p}\ \ \ {\rm by}\ \ \  f_{p}-\frac{\partial\lambda}{\partial x^{p}},
$$
where $\lambda$ is understood to be an arbitrary function of position
in four-space. Here the factor $\frac{e}{ch}$, where $-e$ is the
charge of the electron, $c$ is the speed of light, and
$\frac{h}{2\pi}$ is the quantum of action, has been absorbed in
$f_{p}$. The connection of this ``gauge invariance'' to the
conservation of electric charge remains untouched.
But a fundamental difference, which is important to obtain agreement
with  observation, is that the exponent of the factor
multiplying $\psi$ is not real but pure imaginary. $\psi$ now
plays the role that Einstein's $ds$ played before. It seems to me
that this new principle of gauge-invariance, which follows not
from speculation but from experiment, tells us that the
electromagnetic field is a necessary accompanying phenomenon,
not of gravitation, but of the material wave-field represented
by $\psi$. Since gauge-invariance involves an arbitrary function
$\lambda$ it has the character of ``general'' relativity and can
naturally only be understood in that context.}
\end{quotation}

We shall soon enter into Weyl's justification which is, not
surprisingly, strongly associated with general relativity.
Before this we have to describe his incorporation of the Dirac theory
into GR which he achieved with the help of the tetrad formalism.

One of the reasons for adapting the Dirac theory of the spinning
electron to gravitation had to do with Einstein's recent
unified theory which invoked a distant parallelism with torsion.
E.Wigner \cite{18wig} and others had noticed a connection between this theory
and the spin theory of the electron. Weyl did not like this and
wanted to dispense with teleparallelism. In the introduction
he says:

\begin{quotation}
{\sl I prefer not to believe in distant parallelism for a number
of reasons. First my mathematical intuition objects to accepting
such an artificial geometry; I find it difficult to understand
the force that would keep the local tetrads at different points
and in rotated positions in a rigid relationship. There are,
I believe, two important physical reasons as well. The loosening
of the rigid relationship between the tetrads at different points
converts the gauge-factor $e^{i\lambda}$, which remains arbitrary
with respect to $\psi$, from a constant to an arbitrary function
of space-time. In other words, only through the loosening the
rigidity does the established gauge-invariance become
understandable. }
\end{quotation}

This thought is carried out in detail after Weyl has set up his
two-component theory in special relativity, including a discussion
of $P$ and $T$ invariance. He emphasizes thereby that the
two-component theory excludes a linear implementation of parity
and remarks: ``It is only the fact that the left-right symmetry
actually appears in Nature that forces us to introduce a second pair
of $\psi$-components.'' To Weyl the mass-problem is thus not
relevant for this\footnote{At the time it was thought by Weyl, and
indeed by all physicists, that the 2-component theory requires
a zero mass. In 1957, after the discovery of parity nonconservation,
it was found that the 2-component theory could be consistent with a finite
mass. See K.M. Case, \protect\cite{case}.}. Indeed he says: ``Mass, however, is a
gravitational effect; thus there is hope of finding a substitute
in the theory of gravitation that would produce the required
corrections.''

\subsection{Tetrad Formalism}


In order to incorporate his two-component spinors into GR, Weyl was forced
to make use of local tetrads (Vierbeine). In section 2 of his paper he
develops the tetrad formalism in a systematic manner. This was presumably
independent work, since he does not give any reference to other authors.
It was, however, mainly E.~Cartan who demonstrated with his work~\cite{17n} the
usefulness of locally defined orthonormal bases --also called moving
frames-- for the study of Riemannian geometry.

In the tetrad formalism the metric is described by an
arbitrary  basis of
orthonormal vector fields $\{e_{\alpha}(x);\alpha=0,1,2,3\}$.
If $\{e^{\alpha}(x)\}$ denotes the dual basis of 1-forms,
the metric is given by
\be                                                  \label{3.1}
g=\eta_{\mu\nu}e^{\mu}(x)\otimes e^{\nu}(x),\ \ \ \
(\eta_{\mu\nu})=diag(1,-1,-1,-1).
\ee
Weyl emphasizes, of course, that only a class of such local tetrads is
determined by the metric: the metric is not changed if the
tetrad fields are subject to spacetime-dependent Lorentz
transformations:

\be                                                  \label{3.2}
e^{\alpha}(x)\rightarrow\Lambda^{\alpha}_{\ \beta}(x)e^{\beta}(x).
\ee
With respect to a tetrad, the connection forms
$\omega=(\omega^{\alpha}_{\ \beta})$ have values in the Lie
algebra of the homogeneous Lorentz group:
\be                                                \label{3.3}
\omega_{\alpha\beta}+\omega_{\beta\alpha}=0.
\ee
(Indices are raised and lowered with $\eta^{\alpha\beta}$
and $\eta_{\alpha\beta}$, respectively.) They are determined
(in terms of the tetrad) by the first structure equation
of Cartan:
\be                                                \label{3.4}
de^{\alpha}+\omega^{\alpha}_{\ \beta}\wedge e^{\beta}=0.
\ee
(For a textbook derivation see, e.g., \cite{18})
Under local Lorentz transformations (\ref{3.2}) the connection
forms transform in the same way as the gauge potential of a non-Abelian gauge
theory:
\be                                                \label{3.5}
\omega(x)\rightarrow \Lambda(x)\omega(x)\Lambda^{-1}(x)-
d\Lambda(x)\Lambda^{-1}(x).
\ee
The curvature forms $\Omega=(\Omega^{\mu}_{\ \nu})$ are
obtained from $\omega$ in exactly the same way as the Yang-Mills field
strength from the gauge potential:
\be                                                \label{3.6}
\Omega=d\omega+\omega\wedge\omega
\ee
(second structure equation).

For a vector field $V$, with components $V^{\alpha}$ relative to
$\{e_{\alpha}\}$, the covariant derivative $DV$ is given by
\be                                                \label{3.7}
DV^{\alpha}=dV^{\alpha}+\omega^{\alpha}_{\ \beta}V^{\beta}.
\ee
Weyl generalizes this in a unique manner to spinor fields $\psi$:
\be                                                \label{3.8}
D\psi=d\psi+\frac{1}{4}\omega_{\alpha\beta}\sigma^{\alpha\beta}\psi.
\ee
Here, the $\sigma^{\alpha\beta}$ describe infinitesimal Lorentz
transformations (in the representation of $\psi$). For a Dirac
field these are the familiar matrices
\be                                               \label{3.9}
\sigma^{\alpha\beta}=\frac{1}{2}[\gamma^{\alpha},\gamma^{\beta}].
\ee
(For 2-component Weyl fields one has similar expressions in terms of
the Pauli matrices.)

With these tools the action principle for the coupled Einstein-Dirac
system can be set up. In the massless case the Lagrangian is
\be                                                   \label{3.10}
{\cal L}=\frac{1}{16\pi G}R-i\bar{\psi}\gamma^{\mu}D_{\mu}\psi,
\ee
where the first term is just the Einstein-Hilbert Lagrangian
(which is linear in $\Omega$). Weyl discusses, of course,
immediately the consequences of the following two symmetries:

(i) local Lorentz invariance,

(ii) general coordinate invariance.

\subsection{The New Form of the Gauge-Principle}

All this is a kind of a preparation for the final section of Weyl's paper,
which has the title ``electric field''. Weyl says:
\begin{quotation}
{\sl We come now to the critical part of the theory.
In my opinion the origin and necessity for the electromagnetic field
is in the following. The components $\psi_{1}$ $\psi_{2}$ are,
in fact, not uniquely determined by the tetrad but only to the
extent that they can still be multiplied by an arbitrary
``gauge-factor'' $e^{i\lambda}$. The transformation of the $\psi$
induced by a rotation of the tetrad is determined only up to such
a factor. In special relativity one must regard this gauge-factor
as a constant because here we have only a single
point-independent tetrad. Not so in general relativity;
every point has its own tetrad and hence its own arbitrary
gauge-factor; because by the removal of the rigid connection between
tetrads at different points  the gauge-factor necessarily
becomes an arbitrary function of position.}
\end{quotation}

In this manner Weyl arrives at the gauge-principle in its modern
 form and emphasizes:
``From the arbitrariness of the gauge-factor in $\psi$
appears the necessity of introducing the electromagnetic potential.''
The first term $d\psi$ in (\ref{3.8}) has now to be replaced
by the covariant gauge derivative $(d-ieA)\psi$ and the
nonintegrable scale factor (\ref{2.1}) of the old theory
is now replaced by a phase factor:
$$
\exp\left(-\int_{\gamma}A\right)\rightarrow
\exp\left(-i\int_{\gamma}A\right),
$$
which corresponds to the replacement of the original gauge
group {\bf R} by the compact group $U(1)$. Accordingly, the original
Gedankenexperiment of Einstein translates now to the
Aharonov-Bohm effect, as was first pointed out by
C.N.~Yang in~\cite{22}. The close connection between gauge
invariance and conservation of charge is again uncovered.
The current conservation follows, as in the original theory,
in two independent ways: On the one hand it is a consequence 
of the field equations for matter plus gauge invariance,
at the same time, however, also of the field equations
for the electromagnetic field plus gauge invariance.
This corresponds to an identity in the coupled system
of field equations which has to exist as a result of gauge invariance.
All this is nowadays familiar to students of physics and does not
need to be explained in more detail.

Much of Weyl's paper penetrated also into his classic book
``The Theory of Groups and Quantum Mechanics'' \cite{19}.
There he mentions also the transformation of his early
gauge-theoretic ideas: ``This principle of gauge invariance
is quite analogous to that previously set up by the author,
on speculative grounds, in order to arrive at a unified theory
of gravitation and electricity. But I now believe that this gauge
invariance does not tie together electricity and gravitation,
but rather electricity and matter.''

When Pauli saw the full version of Weyl's paper he became more
friendly and wrote \cite{20}:
\begin{quotation}
{\sl In contrast to the nasty things I said, the essential
part of my last letter has since been overtaken, particularly
by your paper in Z. f. Physik. For this reason I have afterward
even regretted that I wrote to you. After studying
your paper I believe
that I have really understood what you wanted to do
(this was not the case in respect of the little note in the
Proc.Nat.Acad.). First let me emphasize that side of the matter
concerning which I am in full agreement with you: your
incorporation of spinor theory into gravitational theory.
I am as dissatisfied as you are with distant parallelism and your
proposal to let the tetrads rotate independently
at different space-points is a true solution.}
\end{quotation}

In brackets Pauli adds:
\begin{quotation}
{\sl Here I must admit your ability in Physics.
Your earlier theory with $g'_{ik}=\lambda g_{ik}$ was pure
mathematics and unphysical. Einstein was justified in
criticizing and scolding. Now the hour of your revenge
has arrived.}
\end{quotation}

Then he remarks in connection with the mass-problem:
\begin{quotation}
{\sl Your method is valid even for the massive {\rm [Dirac]}
case. I thereby come to the other side of the matter, namely
the unsolved difficulties of the Dirac theory (two signs of
$m_{0}$) and the question of the 2-component theory.
In my opinion these problems will not be solved
by gravitation \ldots the gravitational effects will always be
much too small.}
\end{quotation}

Many years later, Weyl summarized this early tortuous
history of gauge theory in an instructive letter \cite{21} to the
Swiss writer and Einstein biographer C.Seelig, which
we reproduce in 
an English translation.
\begin{quotation}
{\sl
The first attempt to develop a unified field theory of gravitation
and electromagnetism dates to my first attempt in 1918, in which I added the
principle of gauge-invariance to that of coordinate invariance. I myself
have long since abandoned this theory in favour of its correct
interpretation: gauge-invariance as a principle that connects
electromagnetism not with gravitation but with the wave-field of the
electron. ---Einstein was against it} [the original theory]
{\sl from the beginning, and this led to many discussions. I thought
that I could answer his concrete objections. In the end he said
``Well, Weyl, let us leave it at that! In such a speculative manner, without
any guiding physical principle, one cannot make Physics.'' Today one could
say that in this respect we have exchanged our points of view. Einstein 
believes that in this field} [Gravitation and Electromagnetism] {\sl the
gap between ideas and experience is so wide that only the path of
mathematical speculation, whose consequences must, of course, be
developed and confronted with experiment, has a chance of success. Meanwhile
my own confidence in pure speculation has diminished, and I see a need for
a closer connection with quantum-physics experiments, since in my opinion it
is not sufficient to unify Electromagnetism and Gravity. The wave-fields
of the electron and whatever other irreducible elementary particles may
appear must also be included.
}

Independently of Weyl, V.~Fock~\cite{24n} also incorporated the Dirac equation
into GR by using the same method. On the other hand, H.~Tetrode~\cite{25n},
E.~Schr\"odinger~\cite{26n} and V.~Bargmann~\cite{27n} reached this goal by starting
with space-time dependent $\gamma$-matrices, satisfying
$\left\lbrace\gamma^{\mu},\,\gamma^{\nu}\right\rbrace=2\,g^{\mu\nu}$. A
somewhat later work by L.~Infeld and B.L.~van der Waerden~\cite{28n} is
based on spinor analysis.
\end{quotation} 

\section{The Early Work of Kaluza and Klein}

Early in 1919 Einstein received a paper of Theodor Kaluza, a young
mathematician (Privatdozent) and consummate linguist in K\"onigsberg.
Inspired by the work of Weyl one year earlier, he proposed another
geometrical unification of gravitation and electromagnetism by
extending space-time to a five-dimensional pseudo-Riemannian manifold.
Einstein reacted very positively. On 21 April (1919) he writes
``The idea of achieving [a unified theory] by means of a five-dimensional
cylinder world never dawned on me \ldots . At first glance
I like your idea enormously''.  A few weeks later he adds:
``The formal unity of your theory is startling''.
For unknown reasons, Einstein submitted Kaluza's paper to the
Prussian Academy after a delay of two years \cite{[22]}.

Kaluza was actually not the first who envisaged a five-dimensional
unification. It is astonishing that G. Nordstr\"om had this idea
already in 1914 \cite{[23]}. We recall that Nordstr\"om  had worked out
in several papers \cite{[24]} a scalar theory of gravitation that was regarded
by Einstein as the only serious competitor to GR\footnote{For instance,
Einstein extensively discussed Nordstr\"om's second version in his famous
Vienna lecture ``On the Foundations of the Problem of Gravitation"
(23 September 1913), and made it clear that Nordstr\"om's theory was
a viable alternative of his own attempt with Grossmann.
(See Doc.~17 of Vol.~4 of the Collected Papers by
Einstein~\protect\cite{vol8}).}
(In collaboration with Fokker, Einstein gave this theory a
generally covariant (conformally flat) form.)
Nordstr\"om started in his unification attempt with
five-dimensional electrodynamics and imposed the
``cylinder condition", that the fields should not depend
on the fifth coordinate. Then the five-dimensional
gauge potential $^{(5)}\!A$ splits as $^{(5)}\!A=A+\phi dx^5$,
where $A$ is a four-dimensional gauge potential and $\phi$ is
a space-time scalar field. The Maxwell field splits
correspondingly,  $^{(5)}\!F=F+d\phi\wedge dx^5$,
and hence the free Maxwell Lagrangian becomes
\be
-\frac14 (^{(5)}\!F|^{(5)}\!F)=-\frac14 (F|F)+\frac12 (d\phi|d\phi).
\ee
In this manner Nordstr\"om arrived at a unification of his theory
of gravity and electromagnetism. [The matter source
(five-current) is decomposed correspondingly.]
It seems that this early attempt left, as far as we know, no traces
in the literature.

Back to Kaluza's attempt. Like Nordstr\"om he assumes
the cylinder condition. Then the five-dimensional
metric tensor splits into the four-dimensional fields
$g_{\mu\nu}$, $A_\mu$, and $\phi$.
Kaluza's identification of the electromagnetic potential
is not quite the right one, because he chooses it equal to
$g_{\mu 5}$ (up to a constant), instead of
taking the quotient $g_{\mu 5}/g_{55}$.
This does not matter in his further analysis,
because he considers only the linearized approximation
of the field equations. Furthermore, the matter part
is only studied in a non-relativistic approximation.
In particular, the five-dimensional geodesic equation
is only written in this limit. Then the scalar contribution
to the four-force becomes negligible and an automatic split
into the usual gravitational and electromagnetic parts is obtained.

Kaluza was aware of the limitations of his analysis, but he was
confident of being on the right track, as becomes evident
from the final paragraph of his paper:

\bq
{\sl In spite of all the physical and theoretical difficulties
which are encountered in the above proposal it is hard to believe
that the derived relationships, which could hardly be surpassed
at the formal level, represent nothing more than a malicious
coincidence. Should it sometimes be established that the scheme
is more that an empty formalism this would signify a new
triumph for Einstein's General Theory of Relativity,
whose suitable extension to five dimensions
is our present concern.}
\eq

For good reasons the role of the scalar field was unclear to him.

In the classical part of his first paper \cite{[25]},
Klein improves on Kaluza's treatment.  He assumes, however,
beside the condition of cylindricity, that $g_{55}$ is a constant.
Following Kaluza, we keep here the scalar field $\phi$ and
write the Kaluza-Klein ansatz for the five-dimensional metric
$^{(5)}\!g$ in the form
\be                                                  \label{1}
^{(5)}\!g=\phi^{-1/3}(g-\phi\,\omega\otimes\omega)    ,
\ee
where $g=g_{\mu\nu}dx^\mu dx^\nu$ is the space-time metric
and $\omega$ is a differential 1-form of the type
\be                                                \label{2}
\omega=dx^5+\kappa A_\mu dx^\mu.
\ee
Like $\phi$, $A=A_\mu dx^\mu$ is independent of $x^5$;
$\kappa$ is a coupling constant to be determined.
The convenience of the conformal factor $\phi^{-1/3}$
will become clear shortly.

Klein considers the subgroup of five-dimensional
coordinate transformations which respect the form (\ref{1}) of
the $d=5$ metric:
\be                                              \label{3}
x^\mu\rightarrow x^\mu,\ \ \ \  x^5\rightarrow x^5+f(x^\mu).
\ee
Indeed, the pull-back of $^{(5)}\!g$ is again of the form (\ref{1})
with
\be
g\rightarrow g,\ \ \
\phi\rightarrow\phi,\ \ \
A\rightarrow A+\frac{1}{\kappa}df.
\ee
Thus,  $A=A_\mu dx^\mu$  transforms like a gauge potential
under the Abelian gauge group (\ref{3}) and is therefore
interpreted as the electromagnetic potential. This is further
justified by the most remarkable result derived by Kaluza and Klein,
often called Kaluza-Klein miracle. It turns out that the
five-dimensional Ricci scalar $^{(5)}\!R$ splits as follows
\be
^{(5)}\!R=\phi^{1/3}\left(R+\frac14\,\kappa\phi F_{\mu\nu}F^{\mu\nu}
-\frac{1}{6\phi^2}(\nabla\phi)^2+\frac13\,\Delta\ln\phi\right) .
\ee
For $\phi\equiv 1$ this becomes the Lagrangian of the coupled
Einstein-Maxwell system. In view of the gauge group (\ref{3}) this
split is actually no miracle, because no other gauge invariant
quantities can be formed.

For the development of gauge theory this dimensional reduction
was particularly important, because it revealed a close connection
between coordinate transformations in higher-dimensional spaces
and gauge transformations in space-time.

With Klein we consider the $d=5$ Einstein-Hilbert action
\be
^{(5)}\!S=\frac{1}{\kappa^2 L}\int\,  ^{(5)}\!R\sqrt{| ^{(5)}\!g|}d^5 x,
\ee
assuming that the higher-dimensional space is a cylinder
with $0\leq x^5\leq L=2\pi R_5$. Since
\be
\sqrt{| ^{(5)}\!g|}\,dx^5=\sqrt{-g}\,\phi^{-1/3}d^4x\, dx^5
\ee
we obtain
\be                                                  \label{8}
^{(5)}\!S=\int\left(\frac{1}{\kappa^2}\, R
+\frac{1}{4}\,\phi F_{\mu\nu}F^{\mu\nu}
-\frac{1}{6\kappa^2\phi^2}\,(\nabla\phi)^2\right)\sqrt{-g}\,d^4 x.
\ee
Our choice of the conformal factor $\phi^{-1/3}$ in (\ref{1})
was made such that the gravitational part in (\ref{8}) is just
the Einstein-Hilbert action, if we choose
\be
\kappa^2=16\pi G.
\ee
For $\phi\equiv 1$ a beautiful geometrical unification
of gravitation and electromagnetism is obtained.

We pause by noting that nobody in the early history
of Kaluza-Klein theory seems to have noticed the following
inconsistency in putting $\phi\equiv 1$
(see, however, \cite{[28]}):
The field equations for the dimensionally reduced
action (\ref{8}) are just the five-dimensional
equations $^{(5)}\!R_{ab}=0$ for the Kaluza-Klein ansatz (\ref{1}).
Among these, the $\phi$-equation, which is equivalent to
$^{(5)}\!R_{55}=0$, becomes
\be
\Box (\ln\phi)=\frac34\,\kappa^2\phi F_{\mu\nu}F^{\mu\nu}.
\ee
For $\phi\equiv 1$ this implies the unphysical result
$F_{\mu\nu}F^{\mu\nu}=0$. This conclusion is avoided
if one proceeds in the reversed order, i.e., by putting $\phi\equiv 1$
in the action (\ref{8}) and varying afterwards.
However, if the extra dimension is treated as physical --
a viewpoint adopted by Klein (as we shall see) --
it is clearly essential that one maintains consistency
with the $d=5$ field equations. This is an example
for the crucial importance of scalar fields
in Kaluza-Klein theories.

Kaluza and Klein both studied the $d=5$ geodesic equation.
For the metric (\ref{1}) this is just the Euler-Lagrange
equation for the Lagrangian
\be
L=\frac12\, g_{\mu\nu}\dot{x}^{\mu}\dot{x}^{\nu}
-\frac12\,\phi(\dot{x}^{5}+\kappa A_{\mu}\dot{x}^{\mu})^2.
\ee
Since $x^5$ is cyclic, we have the conservation law
($m$=mass of the particle)
\be
p_5/m:=\frac{\partial L}{\partial \dot{x}^{5}}=
\phi(\dot{x}^{5}+\kappa A_{\mu}\dot{x}^{\mu})={\rm const.}
\ee
If use of this is made in the other equations, we obtain
\be                                                    \label{13}
\ddot{x}^{\mu}+\Gamma^{\mu}_{\alpha\beta}\dot{x}^{\alpha}\dot{x}^{\beta} =
-\frac{p_5}{m}\,\kappa\, F^{\mu}_{\ \nu}\,\dot{x}^{\nu}
-\frac12\left(\frac{p_5}{m}\right)^2\phi^{-2}\nabla^\mu\phi.
\ee
Clearly, $p_5$ has to be interpreted as $q/\kappa$, where
$q$ is the charge of the particle,
\be                                             \label{14}
p_5=q/\kappa.
\ee

The physical significance of the last term in (\ref{13})
remained obscure. Much later, Jordan \cite{[26]} and Thiry \cite{[27]}
tried to make use of the new scalar field in order to obtain a theory
in which the gravitational constant is replaced by a dynamical field.
Further work by Fierz \cite{[29]}, Jordan \cite{[26]}, Brans and Dicke 
\cite{[30]}
led to a much studied theory which has been for many years
a serious competitor of GR. Generalized versions have recently
played a role in models of inflation \cite{[31]}.
The question whether the low-energy effective theory
of string theories, say, has Brans-Dicke type interactions,
has lately been investigated for instance by T.Damour
and collaborators \cite{[32]}.

Since the work of Fierz (published in German \cite{[29]})
is not widely known, we briefly describe its main point.
Quoting Pauli \cite{[33]}, Fierz emphasizes, that in theories
containing both a tensor and a scalar fields, the
tensor field appearing most naturally in the action of the theory
can differ from the `physical' metric by some conformal factor
depending on the scalar fields. In order to decide which is the
`atomic-unit' metric and thus the gravitational constant,
one has to look at the coupling to matter.
The `physical' metric $g_{\mu\nu}$ is the one to which matter
is universally coupled (in accordance with the principle of
equivalence).  For instance, the action for a spin-0 massive matter field
$\psi$ should take the form
\be
S_\psi=\frac12\int(g^{\mu\nu}\partial_\mu\psi\partial_\nu\psi
-m^2\psi^2)\sqrt{-g}\,d^4 x.
\ee
A unit of length is then provided by the Compton wave length $1/m$,
and test particles fall along geodesics of $g_{\mu\nu}$.
Fierz generalized the Jordan theory to what is now called
the Brans-Dicke theory. He did not, however, confront
the theory with observations, because he did not believe
in its physical relevance. (The intention of Fierz's
publication was mainly pedagogical (private communication).)

The relation (\ref{14}) brings us to the part
of Klein's first paper that is related to quantum theory.
There he interprets the five-dimensional geodesic equation
as the geometrical optical limit of the wave equation
$^{(5)}\!\Box\Psi=0$ on the higher dimensional space
and establishes for special situations a close relation
of the dimensionally reduced wave equation with Schr\"odinger's equation
that had been discovered in the same year. His ideas
are more clearly spelled out shortly afterwards
in a brief Nature note that has the title
``The Atomicity of Electricity as a Quantum Theory Law" \cite{[34]}.
There Klein says in connection with  Eq.(\ref{14}):

\bq
{\sl The charge q, so far as our knowledge goes,
is always a multiple of the electronic charge $e$, so that
we may write
\be                                            \label{15}
p_5=n\,\frac{e}{\kappa}\ \ \ \ \ [n\in {\rm Z}].
\ee
This formula suggests that the atomicity of electricity
may be interpreted as a quantum theory law.
In fact, if the five-dimensional space is assumed to be closed
in the direction of $x^5$ with period $L$, and if we apply
the formalism of quantum mechanics to our geodesics,
we shall expect $p_5$ to be governed by the following rule:
\be                                             \label{16}
p_5=n\,\frac{h}{\kappa},
\ee
$n$ being a quantum number, which may be positive or negative
according to the sense of motion in the direction of the fifth
dimension, and $h$ the constant of Planck.}
\eq
Comparing (\ref{15}) and (\ref{16}) he finds the value of the
period $L$,
\be                                               \label{17}
L=\frac{hc}{e}\sqrt{16\pi G}=0.8\times 10^{-30}\ {\rm cm},
\ee
and adds:

\bq
{\sl The small value of this length together with the periodicity
in the fifth dimension may perhaps be taken as a support of the theory of
Kaluza in the sense that they may explain the non-appearance
of the fifth dimension in ordinary experiments as the result
of averaging over the fifth dimension.}
\eq

Klein concludes this note with the daring speculation that the
fifth dimension might have something to do with Planck's constant:

\bq
{\sl In a former paper the writer has shown that the differential
equation underlying the new quantum mechanics of Schrodinger
can be derived from a wave equation of a five-dimensional space,
in which $h$ does not appear originally, but is introduced
in connection with the periodicity in $x^5$.
Although incomplete, this result, together with the
considerations given here, suggests that the origin
of Planck's quantum may be sought just in this periodicity
in the fifth dimension.}
\eq

This was not the last time that such speculations have been put forward.
The revival of (supersymmetric) Kaluza-Klein theories
in the eighties~\cite{lo3} led to the idea that the compact dimensions
would necessarily give rise to an enormous quantum vacuum energy
via the Casimir effect.  There were attempts to exploit this
vacuum energy in a self-consistent approach to compactification,
with the hope that the size of the extra dimensions would be
{\em calculable} as a pure number times the Planck length.
Consequently the gauge coupling constant would then be calculable.

Coming back to Klein we note that he would also have arrived at
(\ref{16}) by the dimensional reduction of his five-dimensional
equation.   Indeed, if the wave field $\psi(x,x^5)$ is Fourier
decomposed with respect to the periodic fifth coordinate
\be
\psi(x,x^5)=\frac{1}{\sqrt{L}}\sum_{n\in{\rm Z}}
\psi_n(x){\rm e}^{inx^5/R_5},
\ee
one obtains for each amplitude $\psi_n(x)$
(for the metric (\ref{1}) with $\phi\equiv 1$) the following
four-dimensional wave equation
\be
\left(D^\mu D_\mu-\frac{n^2}{R_{5}^2}\right)\psi_n=0,
\ee
where $D_\mu$ is the doubly covariant derivative
(with respect to $g_{\mu\nu}$ and $A_\mu$) with the charge
\be                                             \label{20}
q_n=n\,\frac{\kappa}{R_5}.
\ee
This shows that the mass of the $n$-th mode is
\be                                           \label{21}
m_n=|n|\frac{1}{R_5}.
\ee
Combining (\ref{20})  with $q_n=ne$, we obtain as before
eq.(\ref{17}) or
\be
R_5=\frac{2}{\sqrt{\alpha}}\, l_{\rm Pl},
\ee
where $\alpha$ is the fine structure constant and $l_{\rm Pl}$
is the Planck length.

Eqs.(\ref{20}) and (\ref{21}) imply a serious defect of the
five-dimensional theory: The (bare) masses of all charged
particles ($|n|\geq 1$) are of the order of the Planck mass
\be
m_n=n\frac{\sqrt{\alpha}}{2}\, m_{\rm Pl}.
\ee

The pioneering papers of Kaluza and Klein were taken up by many
authors. For some time the `projective' theories of Veblen \cite{[35]},
Hoffmann \cite{[36]} and Pauli \cite{[40]} played a prominent role.
These are, however, just equivalent formulations of Kaluza's
and Klein's unification of the gravitational and the
electromagnetic field \cite{[37]}.

Einstein's repeated interest in five-dimensional generalizations
of GR has been described by A. Pais \cite{[38]} and P. Bergmann \cite{[37]},
and will not be discussed here.

\section{Klein's 1938 Theory}

The first attempt to go beyond electromagnetism and gravitation and 
apply Weyl's gauge principle to the nuclear forces occurred in a 
remarkable paper by Oskar Klein, presented at the Kazimierz 
Conference on New Theories in Physics~\cite{[39]} in 1938. Assuming that
the mesons proposed by Yukawa were vectorial, Klein proceeded to construct a 
Kaluza-Klein (KK) like theory which would incorporate them. As in 
the original  KK theory he introduced only one extra dimension 
but his theory differed from the original in two respects:

(i) The fields were not assumed to be completely independent of the 
fifth coordinate
$x^5$ but to depend on it through a factor 
$e^{-ie x^5}$ where $e$ is the electric charge. 

(ii)The 5-dimensional metric-tensor was assumed to be of the form 
\begin{equation}g_{\mu\nu}(x)\,, \qquad g_{55}=1\,, \qquad g_{\mu 5}=\beta
\chi_\mu(x)\,,
\label{s51}\end{equation}
where $g_{\mu\nu}$ was the usual 4-dimensional Einstein metric, 
$\beta$ was 
a constant and $\chi_\mu(x)$ was a {\it matrix-valued} field of the form
\begin{equation}\chi_\mu(x)=\pmatrix{A_\mu(x)& \tilde B_\mu(x)\cr B_\mu(x) & 
A_\mu(x)}=\sigma_3\bigl(\vec\sigma\cdot\vec
A_\mu(x)\bigr)\,,\label{s52}\end{equation}
where the $\sigma$'s are the usual Pauli matrices and $\vec 
A_\mu(x)$ is what we would nowadays call an $SU(2)$ gauge-potential. 
This was a most remarkable Ansatz considering that it implies a 
matrix-valued metric and it is not clear what motivated Klein to 
make it. The reason that he multiplied the present-day $SU(2)$ matrix 
by $\sigma_3$ is that $\sigma_3$ represented the charge matrix for 
the fields. 

Having made this Ansatz Klein proceeded in the standard KK manner 
and obtained instead of the Einstein Maxwell equations a set of 
equations that we would now call the Einstein-Yang-Mills equations. 
This is a little surprising because Klein inserted 
only electromagnetic gauge-invariance. However one can 
see how the $U(1)$ gauge-invariance of electromagnetism could 
generalize to $SU(2)$ gauge-invariance by considering the field 
strengths. The $SU(2)$ form of the field-strengths corresponding to 
the $\tilde B_\mu$ and $B_\mu$ fields, namely
\begin{equation}
F^{\tilde B}_{\mu\nu}=\partial_\mu \tilde B_\nu-\partial_\nu \tilde B_\mu 
+ie(A_\mu \tilde B_\nu-A_\nu \tilde B_\mu)\,,
\label{s53}\end{equation}
\begin{equation}
F^{B}_{\mu\nu}=\partial_\mu B_\nu-\partial_\nu B_\mu
-ie(A_\mu B_\nu-A_\nu B_\mu)\,,
\label{s54}\end{equation}
actually follows from the electromagnetic gauge principle 
$\partial_\mu\rightarrow D_\mu=\partial_\mu+ie(1-\sigma_3)A_\mu$, 
given that the three vector fields belong to the same $2\times 2$ 
matrix. The more difficult question is why the expression 
\begin{equation}F^A_{\mu\nu}=\partial_\mu A_\nu-\partial_\nu A_\mu 
-ie(\tilde B_\mu B_\nu-\tilde B_\nu B_\mu)\label{s55}\end{equation}
for the field strength corresponding to $A_\mu$ contained a bilinear 
term when most other vector-field theories, such as the Proca 
theory, contained only the linear term. The reason is that 
geometrical nature of the dimensional reduction meant that the usual 
space-time derivative $\partial_\mu$ is replaced by the covariant 
space-time derivative $\partial_\mu +ie(1-\sigma_3)\chi_\mu/2$, with 
the result that the usual curl $\partial\wedge\chi $ is replaced by 
$\partial_\mu\chi_\nu-\partial_\nu\chi_\mu+ie/2[\chi_\mu,\chi_\nu]$,
whose third component is just the expression for $F^A_{\mu\nu}$. 

Being interested primarily in the application of his theory to 
nuclear physics, Klein immediately introduced the nucleons, treating 
them as an isodoublet $\psi(x)$ on which the matrix $\xi_\mu$ acted by 
multiplication. In this way he was led to field equations of the 
familiar $SU(2)$ form, namely 
\begin{equation}(\gamma\cdot D+M)\psi(x)=0\,,  \qquad D_\mu=\partial_\mu 
+{ie \over 2}(1-\sigma_3)\chi_\mu\,.\label{s56}\end{equation}
However, although the equations of motion for the vector fields 
$A_\mu$ $\tilde B_\mu$ and $B_\mu$ would be immediately recognized 
today as those of an $SU(2)$ gauge-invariant theory, this was not at 
all obvious at the time and Klein does not seem to have 
been aware of it. Indeed, he immediately proceeded to break the 
$SU(2)$ gauge-invariance by assigning ad hoc mass-terms to the 
$\tilde B_\mu $ and $B_\mu$ fields. 

An obvious weakness of Klein's theory is that there is only one 
coupling constant $\beta$, which implies that the nuclear 
and electromagnetic forces would be of approximately the same 
strength, in contradiction with experiment. 
Furthermore, the nuclear forces would not be charge independent, as 
they were known to be at the time. These weaknesses were 
noticed by M{\o}ller, who, at the end of the talk, objected to the 
theory on these grounds. Klein's answer to these objections was 
astonishing: this problem could easily be solved he said, because 
the strong interactions could be made charge independent (and  
the electromagnetic field seperated) by introducing one more vector field 
$C_\mu$ and generalizing the $2\times 2$ matrix $\chi_\mu$  
\begin{equation}\hbox{from} \quad \chi_\mu=\sigma_3(\vec\sigma\cdot\vec A_\mu) \quad 
\hbox{to} \quad 
\sigma_3(C_\mu+\vec\sigma\cdot\vec A_\mu)\,.\label{s57}\end{equation}
In other words, he there and then generalized what was effectively 
a (broken) $SU(2)$ gauge theory to a broken $SU(2)\times U(1)$ gauge theory. 
In this way, he anticipated the mathematical 
structure of the standard electroweak theory by twenty-one years!

Klein has certainly not forgotten his ambitious proposal of 1938, as
has been suspected in~\cite{50n}. In his invited lecture at the Berne Congress
in 1955~\cite{51n} he came back to some main aspects of his early attempt, and
concluded with the statement:
\begin{quotation}
{\sl
On the whole, the relation of the theory to the fivedimensional
representation of gravitation and electromagnetism on the one hand and to
symmetric meson theory on the other hand --through the appearance of the
charge invariance group-- may perhaps justify the confidence in its
essential soundness.
}
\end{quotation}

\section{The Pauli Letters to Pais}

The next attempt to write down a gauge-theory for the nuclear interactions
was due to Pauli.
During a discussion following a talk by Pais at the 1953
Lorentz Conference in Leiden \cite{24}, Pauli said:

\bq
{\sl\ldots\ I would like to ask in this connection whether the
transformation group with constant phases can be amplified in way
analogous to the gauge group for electromagnetic potentials
in such a way that the meson-nucleon interaction is connected
with the amplified group \ldots}
\eq

Stimulated by this discussion, Pauli worked on this problem and
drafted a manuscript to Pais that begins with~\cite{8}:

\bq
{\sl
Written down July 22--25 1953, in order to see how it looks.
Meson-Nucleon Interaction and Differential Geometry.}
\eq

In this manuscript Pauli generalizes
in this manuscript the Kaluza-Klein theory to a sixdimensional space,
and arrives through dimensional reduction at the essentials
of an $SU(2)$ gauge theory. The extra-dimensions form a two-sphere
$S^2$ with space-time dependent metrics on which $SU(2)$ operates in a
spacetime dependent manner. Pauli develops first in ``local language"
the geometry of what we now call a fiber bundle with a homogeneous
space as typical fiber (in his case $S^2\cong SU(2)/U(1)$).
Studying the curvature of the higher dimensional space, Pauli
automatically finds for the first time the correct expression
for the non-Abelian field strength.

Since it is somewhat difficult to understand what Pauli exactly did,
we give some details, using more familiar formulations and notations.

Pauli considers the sixdimensional total space $M\times S^2$,  where
$S^2$ is the two-sphere on which SO(3) acts in the canonical manner.
He distinguishes among the diffeomorphisms (coordinate transformations)
those which leave $M$ pointwise fixed and induce space-time dependent
rotations on $S^2$:
\be                                                     \label{6.1}
(x,y)\rightarrow (x,R(x)\cdot y).
\ee
Then Pauli postulates a metric on $M\times S^2$ which is supposed
to satisfy three assumptions. These lead him to what is now called
the non-Abelian KK Ansatz: The metric $\hat{g}$ on the total space
is constructed from a space-time metric $g$, the standard metric
$\gamma$ on $S^2$ and a Lie-algebra-valued 1-form
\be
A=A^a T_a,\ \ \ A^a=A^{a}_{\mu}dx^\mu
\ee
on $M$ ($T_a$, $a=1,2,3$, are the standard generators of the
Lie algebra of SO(3)) as follows: If $K^{i}_{a}\partial/\partial y^i$
are the three Killing fields on $S^2$, then
\be
\hat{g}=g-\gamma_{ij}(dy^i+K^{i}_{a}(y)A^a)\otimes(dy^j+K^{j}_{a}(y)A^a).
\ee
In particular, the non-diagonal metric components are
\be                                                \label{6.4}
\hat{g}_{\mu i}=A^{a}_{\mu}(x)\gamma_{ij}K^{j}_{a}.
\ee
Pauli does not say that the coefficients of $A^{a}_{\mu}$ in (\ref{6.4})
are the components of the three independent Killing fields.
This is, however, his result which he formulates in terms of
homogeneous coordinates for $S^2$. He determines the transformation
behavior of $A^{a}_{\mu}$ under the group (\ref{6.1}) and finds
in matrix notation what he calls ``the generalization of the gauge group":
\be
A_{\mu}\rightarrow RA_{\mu}R^{-1}+R^{-1}\partial_\mu R.
\ee

With the help of $A_\mu$ he defines a covariant derivative
which is used to derive ``field strengths" by applying a
generalized curl to $A_\mu$. This is exactly the field strength
which was later introduced by Yang and Mills. To our knowledge,
apart from Klein's 1938 paper, it appears here the first time. 
Pauli says that ``this is the
{\em true} physical field, the analog of the {\em field strength}"
and he formulates what he considers to be his ``main result":

\bq
{\sl 
The vanishing of the field strength is necessary and sufficient
for the $A^{a}_\mu(x)$ in the whole space to be transformable to zero.}
\eq

It is somewhat astonishing that Pauli did not work out the Ricci scalar
for $\hat{g}$ as for the KK-theory. One reason may be connected
with his remark on the KK-theory in Note 23 of his relativity
article~\cite{15} concerning the five-dimensional curvature scalar (p.230):

\bq
{\sl There is, however, no justification for the particular choice
of the five-dimensional curvature scalar $P$ as integrand of the action
integral, from the standpoint of the restricted group of the
cylindrical metric [gauge group]. The open problem of finding
such a justification seems to point to an amplification
of the transformation group.}
\eq

In a second letter (reprinted in \cite{6}), Pauli also studies
the dimensionally reduced Dirac equation and arrives at a
mass operator which is closely related to the Dirac operator
in internal space $(S^2,\gamma)$.  The eigenvalues
of the latter operator had been determined by him
long before~\cite{[41]}. Pauli concludes with the statement:
``So this leads to some rather unphysical `shadow particles'."

\section{Yang-Mills Theory}

In his Hermann Weyl Centenary Lecture at the ETH \cite{22},
C.N. Yang commented on Weyl's remark ``The principle of gauge-invariance
has the character of general relativity since it contains an arbitrary
function $\lambda$, and can certainly only be understood in terms
of it'' \cite{23} as follows:
\begin{quotation}
{\sl The quote above from Weyl's paper also contains something
which is very revealing, namely, his strong association of gauge
invariance with general relativity. That was, of course, natural since
the idea had originated in the first place with Weyl's attempt in 1918
to unify electromagnetism with gravity. Twenty years later, when
Mills and I worked on non-Abelian gauge fields, our motivation
was completely divorced from general relativity and we did not
appreciate that gauge fields and general relativity are somehow
related. Only in the late 1960's did I recognize the structural
similarity mathematically of non-Abelian gauge fields with general
relativity and understand that they both were connections
mathematically.}
\end{quotation}

Later, in connection with Weyl's strong emphasis of the relation
between gauge invariance and conservation of electric charge,
Yang continues with the following instructive remarks:
\begin{quotation}
{\sl Weyl's reason, it turns out, was also one of the melodies
of gauge theory that had very much appealed to me when as a
graduate student I studied field theory by reading Pauli's articles.
I made a number of unsuccessful attempts to generalize gauge theory
beyond electromagnetism, leading finally in 1954 to a
collaboration with Mills in which we developed a non-Abelian gauge
theory. In [\ldots ] we stated our motivation as follows:

The conservation of isotopic spin points to the existence of a
fundamental invariance law similar to the conservation of electric
charge. In the latter case, the electric charge serves as a source
of electromagnetic field; an important concept in this case is
gauge invariance which is closely connected with
(1) the equation of motion of the electro-magnetic field,
(2) the existence of a current density, and
(3) the possible interactions between a charged field and the
electromagnetic field. We have tried to generalize this concept
of gauge invariance to apply to isotopic spin conservation.
It turns out that a very natural generalization is possible.

Item (2) is the melody referred to above. The other two melodies,
(1) and (3), where what had become pressing in the early 1950's
when so many new particles had been discovered and physicists
had to understand how they interact which each other.

I had met Weyl in 1949 when I went to the Institute for Advanced
Study in Princeton as a young ``member''. I saw him from time
to time in the next years, 1949--1955. He was very approachable, 
but I don't remember having discussed physics or mathematics
with him at any time. His continued interest in the idea of gauge
fields was not known among the physicists. Neither
Oppenheimer nor Pauli ever mentioned it. I suspect they also 
did not tell Weyl of the 1954 papers of Mills' and mine.
Had they done that, or had Weyl somehow came across our paper,
I imagine he would have been pleased and excited, for we had
put together two things that were very close to his heart:
gauge invariance and non-Abelian Lie groups.}
\end{quotation}

It is indeed astonishing that during those late years neither Pauli
nor Yang ever talked with Weyl about non-Abelian generalizations of
gauge-invariance.

With the background of Sect.~6, the following story of spring 1954 becomes
more understandable. In late February, Yang was invited by
Oppenheimer to return to Princeton for a few days and to give
a seminar on his joint work with Mills. Here, Yang's report \cite{25}:
\begin{quotation}
{\sl  Pauli was spending the year in Princeton, and was deeply
interested in symmetries and interactions. (He had written in
German a rough outline of some thoughts, which he had sent to
A. Pais. Years later F.J. Dyson translated this outline into
English. It started with the remark, ``Written down July 22-25,
1953, in order to see how it looks,'' and had the title
``Meson-Nucleon Interaction and Differential Geometry.'')
Soon after my seminar began, when I had written down on the
blackboard,
$$
(\partial_{\mu}-i\epsilon B_{\mu})\psi,
$$
Pauli asked, ``What is the mass of this field $B_{\mu}$?''
I said we did not know. Then I resumed my presentation, but soon
Pauli asked the same question again. I said something to the effect
that that was a very complicated problem, we had worked on it
and had come to no definite conclusions. I still remember his
repartee: ``That is not sufficient excuse.'' I was so taken aback
that I decided, after a few moments' hesitation to sit down.
There was general embarrassment. Finally Oppenheimer said,
``We should let Frank proceed.'' I then resumed, and Pauli did not
ask any more questions during the seminar.

I don't remember what happened at the end of the seminar. But
the next day I found the following message:

February 24, Dear Yang, I regret that you made it almost impossible
for me to talk with you after the seminar. All good wishes.
Sincerely yours, W.Pauli.

I went to talk to Pauli. He said I should look up a paper by
E. Schr\"odinger, in which there were similar mathematics\footnote{E. Schr\"odinger, Sitzungsberichte der Preussischen
(Akademie der Wissenschaften, 1932), p. 105.}.
After I went back to Brookhaven, I looked for the paper and
finally obtained a copy. It was a discussion of
space-time-dependent representations of the $\gamma_{\mu}$
matrices for a Dirac electron in a gravitational field.
Equations in it were, on the one hand, related to equations
in Riemannian geometry and, on the other, similar to the
equations that Mills and I were working on. But it was many
years later when I understood that these were all different
cases of the mathematical theory of connections on fiber bundles.}
\end{quotation}

Later Yang adds:
\begin{quotation}
{\sl I often wondered what he [Pauli] would say about the subject
if he had lived into the sixties and seventies.}
\end{quotation}

At another occasion \cite{22} he remarked:
\begin{quotation}
{\sl I venture to say that if Weyl were to come back today,
he would find that amidst the very exciting, complicated and detailed
developments in both physics and mathematics, there are
fundamental things that he would feel very much at home with.
He had helped to create them.}
\end{quotation}

Having quoted earlier letters from Pauli to Weyl, we add what
Weyl said about Pauli in 1946 \cite{26}:
\begin{quotation}
{\sl The mathematicians feel near to Pauli since he is
distinguished among physicists by his highly developed organ for
mathematics. Even so, he is a physicist; for he has to a high
degree what makes the physicist; the genuine interest in the
experimental facts in all their puzzling complexity.
His accurate, instructive estimate of the relative weight
of relevant experimental facts has been an unfailing guide for
him in his theoretical investigations. Pauli combines in an
exemplary way physical insight and mathematical skill.}
\end{quotation}

To conclude this section, let us emphasize the main differences
of GR and Yang-Mills theories. Mathematically, the $so(1,3)$-valued
connection forms $\omega$ in Sec.~3.1 and the Liealgebra-valued
gauge potential $A$ are on the same footing; they are both
representatives of connections in (principle) fiber bundles
over the spacetime manifold. Eq.(\ref{3.6}) translates into
the formula for the Yang-Mills field strength $F$,
\be                                                  \label{4:1}
F=dA+A\wedge A.
\ee
In GR one has, however, additional geometrical structure, since the
connection derives from a metric, or the tetrad fields
$e^{\alpha}(x)$, through the first structure equation (\ref{3.4}).
Schematically, we have:

\begin{figure}[h]
\begin{center}
\epsfig{file=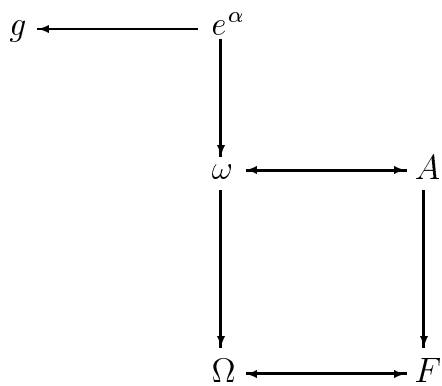,width=6cm}
\caption{General Relativity versus Yang-Mills theory.}
\end{center}
\end{figure}

(In bundle theoretical language one can express this as follows:
The principle bundle of GR, i.e., the orthonormal frame bundle,
is constructed from the base manifold and its metric, and has
therefore additional structure, implying in particular the
existence of a canonical 1-form (soldering form), whose
local representative are the tetrad fields; see, e.g. \cite{Bleecker}.)

Another important difference is that the gravitational Lagrangian
$\ast R=\frac{1}{2}\Omega_{\alpha\beta}\wedge\ast
(e^{\alpha}\wedge e^{\beta})$ is linear in the field strengths,
whereas the Yang-Mills Lagrangian $F\wedge\ast F$ is quadratic.

\section{Recent Developments}

The developments after 1958 consisted in the gradual recognition
that --- contrary to phenomenological appearances --- Yang-Mills gauge theory
can describe weak and strong interactions. This important step was again
very difficult, with many hurdles to overcome.

One of them was the mass problem which was solved, probably in a preliminary
way, through spontaneous symmetry breaking. Of critical significance was
the recognition that spontaneously broken gauge theories are renormalizable.
On the experimental side the discovery and intensive investigation of the
neutral current was, of course, crucial. For the gauge description of
the strong interactions, the discovery of asymptotic freedom was decisive .
That the $SU(3)$ color group should be gauged was also not at all obvious.
And then there was the confinement idea which explains why quarks and gluons
do not exist as free particles. All this is described in numerous modern text
books and does not have to be repeated.

The next step of creating a more unified theory of the basic interactions
will probably be much more difficult. All major theoretical
developments of the last twenty years, such as grand unification,
supergravity
and supersymmetric string theory are almost completely separated from
experience. There is a great danger that theoreticians get lost in pure
speculations. Like in the first unification proposal of Hermann Weyl they may
create beautiful and highly relevant mathematics which does, however, not
describe nature.
In the latter case history shows, however, that such ideas can one day also
become fruitful for physics. It may, therefore, be appropriate to conclude
with some remarks on current attempts in string theory and non-commutative
geometry.

\subsection{Gauge Theory and Strings}
\subsubsection{Introduction}
So far we have considered gravitation and gauge-theory only within the 
context of local field theory. However, gravitation and gauge-theory 
also occur naturally in string theory~\cite{lo1},~\cite{lo2}. 
Indeed, whereas in 
field theory they are optional extras that are introduced on 
phenomenological grounds (equality of inertial and gravitational 
mass, divergence-free character of the magnetic field etc.) 
in string theory they occur as an intrinsic part of the structure. 
Thus string theory is a very natural setting for gravitation and 
gauge-fields. One might go as far as to say that, had string theory 
preceded field theory historically, the gravitational and gauge fields 
might have emerged in a completely different manner. An interesting 
feature of string gauge-theories is that the choice of gauge-group 
is quite limited.  

String theory is actually not only a natural setting for gravitational and 
gauge-fields but also for the Kaluza-Klein (KK) mechanism. 
Historically, 
the most obvious difficulty with  KK reductions was that there is no 
experimental evidence and no theoretical 
need for any extra dimensions. 
String theory changes this situation dramatically. As is well-known, 
string theory is conformally-invariant only if the dimension $d$ of the 
target-space is $d=10$ or $d=26$, according to whether the string is 
supersymmetric or not. Thus, in contrast to field theory,  
string theory points to the existence of extra dimensions and even 
specifies their number. 

We shall treat an important specific case of dimensional reduction 
within string theory, namely the toroidal reduction from 26 to 10 
dimensions, in Sect. 8.1.7. However, since no phenomenologically 
satisfactory reduction from 26 or 10 to 4 dimensions has yet emerged, and the 
dimensional reduction in string theory is rather similar to that 
in ordinary field theory~\cite{lo3} we shall not consider any other case,  
but refer the reader to the literature. 

Instead we shall concentrate on the manner in which 
gauge-theory and gravitation occur in the context of 
dimensionally unreduced string theory. We shall rely heavily on the 
result~\cite{lo4} that if a massless vector field theory with polarization vector 
$\xi_\mu$ and on-shell momentum $p_\mu$ $(p^2=0)$ is invariant with 
respect to the transformation 
\begin{equation}\xi_\mu(p)\rightarrow \xi_\mu(p)+\eta(p)
p_\mu\,,\label{s81}\end{equation}
where $\eta(p)$ is an arbitrary scalar, then it must be a 
gauge theory. Similarly we shall rely on the result that 
a second-rank symmetric tensor theory for which 
the polarization vector $\xi_{\mu\nu}$ satisfies 
\begin{equation}\xi_\mu^\mu(p)=0\,, \qquad p^\mu \xi_{\mu\nu}(p)=0\,,
\label{s82}\end{equation}
and the theory is invariant with respect to 
\begin{equation}\xi_{\mu\nu}(p)\rightarrow \xi_{\mu\nu}(p)+\eta_\mu p_\nu+\eta_\nu 
p_\mu\label{s83}\end{equation}
for arbitrary $\eta_\mu(p)$, and $p^2=0$, must be a gravitational theory. 
A good presentation of these results is given in~\cite{lons}. 

\subsubsection{Gauge Properties of Open Bosonic Strings}

To fix our ideas we first recall the form of the path-integral for the 
bosonic string~\cite{lo1},~\cite{lo2}, namely 
\begin{equation}\int dh dX e^{\int d^2\sigma \sqrt{h}h^{\alpha\beta}\eta_{\mu\nu}
\partial_\alpha X^\mu(\sigma)\partial_\beta
X^\nu(\sigma)}\,,\label{s84}\end{equation}
where $\sigma$, $d^2\sigma$ and $h_{\alpha\beta}$ are the 
coordinates, diffeomorphic-invariant measure and metric on the 
2-dimensional world-sheet of the string respectively and 
$\eta_{\mu\nu}$ is the Lorentz metric for the 26-dimensional 
target-space in which the string, with coordinates $X^\mu(\sigma)$, moves.   
Thus the $X^\mu(\sigma)$ may be regarded as fields in a 
2-dimensional quantum field theory. The action in~(\ref{s84}), and hence the 
classical 2-dimensional field theory, is conformally invariant, but,  
as is well-known, the quantum theory is conformally invariant 
only if $N=26$ or $N=10$ in the supersymmetric version.  

The {\it open} bosonic string is the one in which gauge-fields 
occur. Indeed, one might go as far as to say that the open string 
is a natural non-local generalization of a gauge-field. 
The ends of the open strings are usually assumed to be attached to quarks 
and thus there is a certain qualitative resemblance between the open bosonic 
strings and the gluon flux-lines that link the quarks in 
theories of quark confinement. We wish to make the relationship 
between gauge-fields and open bosonic strings more precise. 

As is well-known~\cite{lo1},~\cite{lo2},  the vacuum state $\vert 0>$ of 
the open string 
is a scalar tachyon and the first excited states are 
$X^\mu_+(\sigma)\vert 0>$, where $X^\mu_+(\sigma)$ denotes the 
positive-frequency parts of the fields $X^\mu(\sigma)$. For a 
suitable, standard, value of the normal-ordering parameter for 
the Noether generators of the conformal (Virasoro) group, these 
states are massless i.e. $p^2=0$, where 
$p_\mu$ is the 26-dimensional momentum. Furthermore, they are the 
only massless states. Thus, if there are gauge fields in the theory, 
these are the states that must be identified with them. On the other 
hand, since all the other (massive) states in the Fock space of the 
string are formed by acting on $\vert 0>$ with higher powers of 
$X^\mu_+(\sigma)$, we see that the operators $X_+^\mu(\sigma)$, 
which create the massless states, create the whole string. Thus the 
open bosonic string can be thought of as a kind of glueball of 
the massless particles, and it is in this sense that it can be 
regarded as a non-local generalization of a gauge-field. The question 
is: how is the identification of the massless states $X_+^\mu(\sigma)\vert 0>$ 
with gauge-fields to be justified? The identification is made through the 
so-called vertex-operators for the emission of the on-shell massless 
states. These vertices are the analogue of the ordinary vertices of 
Feynman diagrams in quantum field theory and take the form 
\begin{equation}V(\xi,p)=\int d^2\sigma e^{ip\cdot X(\sigma)}\partial_s\xi\cdot
X(\sigma)\,,
\label{s85}\end{equation}
where $\xi_\mu$ is the polarization vector, $p_\mu$ is the 
momentum $(p^2=0)$ of the emitted particle and $\partial_s$ is a 
space-like derivative. This 
vertex-operator is to be inserted in the functional integral~(\ref{s84}). 
Although the form of this vertex is not deduced from a second-quantized 
theory of strings (which does not yet exist) 
the vertex-operator ~(\ref{s85}) is generally accepted as the correct one,
because it is the only vertex that is compatible with the 2-dimensional 
structure and conformal invariance of the string, and which reduces to 
the usual vertex in the point-particle limit. Suppose now that we 
make the transformation $\xi_\nu \rightarrow \xi_\nu+\eta(p) p_\nu$ of 
Eq.~(\ref{s81}). Then the vertex $V(\xi,p_\mu)$ acquires the additional term 
\begin{equation}\eta(p) \int d^2\sigma e^{ip\cdot X(\sigma)}\partial_s p\cdot X
=-i\int d^2\sigma\partial_+\Bigl(e^{ip\cdot X(\sigma)}\Bigr) =0\,.
\label{s86}\end{equation}
From this we see that the on-shell vertex-operator is invariant under 
this transformation, which, as discussed earlier, is just the condition 
for the $\xi$ to be the polarization vector of a gauge-field. The 
important point is that this 
gauge-invariance is not imposed from outside, but is an intrinsic 
property of the string. It is a consequence of the fact that 
the string is conformally invariant (which dictates 
the form of the vertex-operator) and has an internal structure (manifested 
by the fact that it has an internal 2-dimensional integration). 

\subsubsection{Gravitational Properties of Closed Bosonic Strings}

Just as the open bosonic string is the one in which gauge-fields 
naturally occur, the closed bosonic string is the one in which 
gravitational fields naturally occur. It turns out, in fact, that a 
gravitational field, a dilaton field and an anti-symmetric tensor field 
occur in the closed string in the same way that the gauge-field appears 
in the open string. The ground state $\vert 0>$ of the closed 
string is again a tachyon but the new feature is that for the standard 
value of the normal-ordering constant, the first excited states 
are massless states of the form $X^\mu_+(\sigma)X^\nu_+(\sigma)\vert 
0>$ and it is the symmetric, trace and anti-symmetric parts of the  
two-tensor formed by the $X$'s that are identified with the gravitational, 
dilaton and anti-symmetric tensor fields, respectively. 
The question is how the identification with the gravitational field 
is to be justified and again the answer is by means of a 
vertex-operator, this time for the emission of an on-shell graviton.  
The vertex-operator that describes the 
emission of an on-shell massless spin-2 field (graviton) of polarization 
$\xi_{\mu\nu}$ and momentum $k_\lambda$, where $k^2=0$, is 
\begin{equation}V_{\mu\nu}=\int d^2\sigma 
e^{ip\cdot X(\sigma)} 
h^{\alpha\beta}\xi_{\mu\nu}\partial_\alpha X^\mu(\sigma)\partial_\beta 
X^\nu(\sigma)\,.\label{s87}\end{equation} 

Already at this stage there is a feature that does not arise in the 
gauge-field case as follows: Since the vertex-operator is bilinear in 
the field $X^\mu$ it has to be 
normal-ordered, and it turns out that the normal ordering 
destroys the classical conformal invariance, unless  
\begin{equation}\xi^\mu_\mu=0  \quad \hbox{and} \quad p^\mu\xi_{\mu\nu}=0\,. 
\label{s88}\end{equation}
We next make the momentum-space version of an infinitesimal coordinate 
transformation, namely
\begin{equation}\xi_{\mu\nu}\rightarrow 
\xi_{\mu\nu}+\eta_\mu(p) p_\nu+\eta_\nu(p) p_\mu\,.
\label{s89}\end{equation} 
Under this 
transformation the vertex $V_\xi$ picks up an additional term of the form 
\begin{eqnarray}
& 2\eta_\nu\int d^2\sigma h^{\alpha\beta}e^{ip\cdot X(\sigma)} 
p_\mu\partial_\alpha X^\mu(\sigma)\partial_\beta X^\nu(\sigma) & \nonumber\\
& =-2i\eta_\nu\int d^2\sigma h^{\alpha\beta}
\Bigl(\partial_\alpha e^{ip\cdot X(\sigma)}\Bigr) 
\partial_\beta X^\nu(\sigma)\,. &
\end{eqnarray}
In analogy with the electromagnetic case we can carry out a partial 
integration. However, this time the expression does not vanish completely 
but reduces to                                            
\begin{equation}2i\eta_\nu\int d^2\sigma \sqrt{h}e^{ik_\lambda X^\lambda(\sigma)}
\Delta X^\nu(\sigma)\,,\label{s89b}\end{equation}
where $\Delta$ denotes the 2-dimensional Laplacian. 
On the other hand, $\Delta X^\nu(\sigma)=0$ is just  
the classical field equation for $X^\nu(\sigma)$, and it can be shown 
that even in the quantized version it is effectively
zero~\cite{lo1},~\cite{lo2}. 
Thus, thanks to the dynamics, we have invariance with respect to the 
transformations~(\ref{s89}).  But~(\ref{s88}) and invariance with respect 
to~(\ref{s89}) are 
just the conditions~(\ref{s82}) and~(\ref{s83}) discussed earlier for the vertex to 
be a gravitational field. As in the gauge-field case, the 
important point is that the general coordinate invariance is not imposed, 
but is a consequence of the conformal invariance and internal structure 
of the string. 

The appearance of a scalar field in this context is not too surprising
since a scalar also appeared in the KK reduction. What is more
surprising is the appearance of an anti-symmetric tensor. 
From the point of view of traditional local gravitational and gauge field 
theory the presence of an additional anti-symmetric tensor field seems at 
first sight to be an embarassment. But it turns out to play an essential 
role in maintaining conformal invariance (cancellation of anomalies), 
so its presence is to be welcomed. 

\subsubsection{The Presence of Matter} 

Of course, the open bosonic string is not the whole story any more than 
pure gauge-fields are the whole story in quantum field theory. One 
still has to introduce quantities that correspond to fermions (and 
possibly scalars) at the mass-zero level. There are essentially two 
ways to do this. The first is the Chan-Paton
mechanism~\cite{lo1},~\cite{lo2}, which 
dates from the days of strong interaction string-theory. In this 
mechanism one simply attaches charged particles to the open ends 
of the string. These charged particles are not otherwise associated 
with any string and thus the mechanism is rather ad hoc and leads to 
a hybrid of string and field theory. But it has the merit of introducing 
charged particles directly and thus emphasizing the relationship between 
strings and gauge fields. 

The Chan-Paton mechanism  has the further merit of allowing a simple 
generalization to the non-Abelian case. This is done by 
replacing the charged particles by particles belonging to the 
fundamental representations of compact internal symmetry groups G,  
typically quarks $q_a(x)$ and anti-quarks $\bar q_{b}(x)$. The 
vertex-operator then generalizes to one with double-labels $(a,b)$ and 
represents non-Abelian gauge-fields in much the same way that the simple 
bosonic string represents an Abelian gauge-field. 

An interesting restriction arises from the fact that since 
the string represents gauge fields and gauge-fields belong to the 
adjoint representation of the gauge-group, the vertex function must 
belong to the adjoint representation. This implies that even at the 
tree-level the tensor product of the fundamental group representation 
with itself must produce only the adjoint representation and this 
restricts $G$ to be one of the classical groups $SO(n)$, $Sp(2n)$ and $U(n)$.  
Furthermore, it is found that $U(n)$ violates unitarity at the one-loop 
level, which leaves only $SO(n)$ and $SP(2n)$. Finally,  
these groups require symmetrization and anti-symmetrization in 
the indices $a$ and $b$ to produce only the adjoint representation 
and this implies that the string be oriented (symmetric with respect 
to its end points). When all these conditions are satisfied it can 
be shown that the non-Abelian vertex 
corresponding to~(\ref{s85}) is covariant respect to the non-Abelian 
gauge-transformations corresponding to 
$\xi_\mu\rightarrow \xi_\mu+\eta(p) p_\mu$ above. 
But since these transformations are non-linear the proof is 
more difficult than in the Abelian case. 

\subsubsection{Fermionic and Heterotic Strings: Supergravity and Non-Abelian Gauge Theory} 

The Chan-Paton version of gauge string has the obvious disadvantage 
that the charged fields (quarks) are not an intrinsic part of the 
theory. 
A second method to introduce fermions is to place them in the string
itself. This is done by replacing the kinetic term $(\frac{\partial}{\partial x})^2$
by a Dirac term $\bar\Psi\partial\hspace{-0.22truecm}\slash\Psi$ in the Lagrangian density.
Interesting cases are those in which the number of fermion components just
matches the number of bosons so that the Lagrangian is supersymmetric. In
that case the condition for quantum conformal invariance reduces from $d=26$
to $d=10$. An interesting case from the point of view of gauge theory and
dimensional reduction is the heterotic string, in which the left-handed part
forms a superstring and the right handed part forms a bosonic string in
which $16$ of the bosons are fermionized.
For the heterotic string the Lagrangian in the bosonic 
string path-integral~(\ref{s84}) is replaced by the Lagrangian
\begin{equation}\sum_{\mu=1}^{\mu=10} \partial_\alpha X^\mu\partial^\alpha X_\mu 
-2\sum_{\mu=1}^{\mu=10}\psi^\mu_+\partial_-\psi_{\mu +}
-2\sum_{A=1}^{A=32}
\lambda_-^A\partial_+\lambda^A_-\,,\label{s810}\end{equation}
where the $\psi$'s and $\lambda$'s are Majorana-Weyl fermions 
and the $\lambda$'s belong to a representation 
(labelled with A) of an internal symmetry group G. 
It is only through the $\lambda$'s that the internal symmetry group 
enters. 
The left and right-handed parts of the theory are conformally invariant for 
quite different reasons. The left-handed part of the $X$'s and the 
left-handed fermions $\psi$ are conformally invariant, because together they 
form the left-handed part of a superstring (this is why the 
summation over the $X$'s is only from 1 up to 10). The right-handed part of 
the $X$'s and the right-handed fermions $\lambda^A$ are anomaly-free 
because, from the point of view of anomalies, two Majorana-Weyl fermions 
are equivalent to one boson and thus the system is equivalent to the 
right-handed part of a 26-dimensional bosonic string. (This is why 
the index A runs from 1 to 32.) The fact that there are 32 fermions 
obviously puts strong restrictions on the choice of the internal 
symmetry group G. 

We now examine the particle content of the theory, using the 
light-cone gauge, where there are no redundant fields. There are no 
tachyons for the left-moving fields, the first excited states are massles 
and take the form 
\begin{equation}\vert i>_L \quad \hbox{and} \quad \vert \alpha>_L\,,
\label{s811}\end{equation}
where the $\vert i>_L$ for $i=1...8$ are the left-handed components 
of a massless space-time vector in the eight transverse directions 
in the light-cone gauge and $\vert\alpha>_L$ are the components of a 
massless fermions in one of the two fundamental spinor representations 
of the same space-time $SO(8)$ group. These states are all G-invariant. 

The first excited states for the right-moving sector are 
\begin{equation}\qquad \vert i>_R \quad \hbox{and}  \qquad  
\lambda^A\lambda^B\vert0>\,, \label{s812}\end{equation} 
where the $\vert i>_R$ are the right-handed analogues of the $\vert 
i>_L$ and the $\lambda\lambda\vert 0>$ states are massless space-time 
scalars. The states $\vert i>_R$ are G-invariant but the states 
$\lambda\lambda\vert0>$ belong to the adjoint representation of $G$ 
and thus it is only through these states that the internal symmetry 
enters at the massless level. 

The physical states are obtained by tensoring the left and 
right-moving states~(\ref{s811}) and~(\ref{s812}). On tensoring the right-handed states 
with the vectors in~(\ref{s812}) we obviously obtain states which are G-invariant, 
and they turn out to be just the states that would occur in $N=1$ supergravity. 
An analysis of the vertex operators, similar to that carried out 
above for closed bosonic strings, confirms that these fields do indeed 
correspond to supergravity. 

\subsubsection{The Internal Symmetry Group G}

From the point of view of non-Abelian gauge-theory the interesting 
states are those belonging to the non-trivial representations of $G$, 
and these are the ones obtained from the tensor products of~(\ref{s812}) with 
the space-time scalars  $\lambda\lambda\vert0>$. At this point 
one must make a choice about the internal symmetry group G. 
The simplest choice is evidently $G=SO(32)$, and it is 
obtained by assigning anti-periodic 
boundary conditions to {\it all} the fermion fields $\lambda$. 
(Assigning periodic boundary conditions to all of them violates 
the masslessness condition.) Since the product states continue to 
belong to the adjoint representation of $SO(32)$, they are the natural 
candidates for states associated with non-Abelian gauge-fields and an 
analyses of the vertex operators associated with these states confirms 
that they do indeed corrrespond to $SO(32)$ gauge-fields.  

In sum, the heterotic string produces both supergravity and non-Abelian 
gauge theory. 

\subsubsection{Dimensional Reduction and the Heterotic Symmetry Group ${\bf 
E_8\times E_8}$}

A variety of other left-handed internal symmetry groups $G\subset 
SO(32)$ can be obtained by assigning periodic and anti-periodic 
boundary  conditions to the fermions $\lambda^A$ of the heterotic 
string in a non-uniform manner. However, apart from the SO(32) case 
just discussed, the only assignment that satisfies unitarity at the 
one-loop level is an equipartition of the 32 fermions into two sets 
of 16, with mixed boundary conditions. This would appear at first 
sight to lead to an $SO(16)\times SO(16)$ internal symmetry and 
gauge group, on the same grounds as $SO(32)$ above, but 
a closer analysis shows that it actually leads to a larger group, 
namely $E_8\times E_8$, which actually has the same dimension (496) as 
$SO(32)$. This group is quite attractive for grand unification theory 
as it breaks naturally to $E_6$, which is one of the favourite 
GUT groups. 

Once we accept that $SO(16)\times SO(16)$ is a gauge-group and 
that a rigid internal symmetry group $E_8\times E_8$ exists, it 
follows immediately that $E_8\times E_8$ must be a gauge-group, 
because the action of the rigid generators of 
$E_8\times E_8$ on the $SO(16)\times SO(16)$ gauge-fields produces  
$E_8\times E_8$ gauge-fields. 

This reduces the problem to the existence of a rigid $E_8\times E_8$ 
symmetry, but, within the context of our present methods, this is a 
rather convoluted process. One must introduce special 
representations of $SO(16)\times SO(16)$, project out some of the 
resulting states and construct vertices that represent 
the elements of the coset $(E_8\times E_8)/(SO(16)\times SO(16))$. Luckily 
there is a much more intuitive way to establish the existence of the  
$E_8\times E_8$ symmetry, and as this way provides a very nice example 
of dimensional reduction within string theory we shall now sketch it. 

We have already remarked that, from the point of view of Virasoro 
anomalies, the 32 right-handed Majorana-Weyl fermions $\lambda$ are 
equivalent to the right-handed parts of 16 bosons. This relationship 
can be carried farther by bosonizing the fermions according to 
$\lambda_{\pm}(\sigma)=:\hbox{exp}\bigl(\pm\phi_i^R(\sigma)\bigr):$, 
where $\phi^R(\sigma)$ is a right-moving bosonic field, compactified 
so that $0\leq \phi(\sigma)<2\pi$. In that case we may 
regard the right-handed part of the heterotic string as originating 
in the right-handed part of an ordinary 26-dimensional bosonic string, 
in which 16 of the 26 right-moving bosonic fields $X^R(\sigma)$ 
have been fermionized by letting $X^R_a(\sigma)\rightarrow \phi^R_a(\sigma)$ 
for $0\leq\phi_a(\sigma)<2\pi$ and $a=1...16$. Since the $X$'s 
correspond to coordinates in the target space of the string, this 
is equivalent to a toroidal compactification of 16 of 
dimensions of the target space and thus is equivalent of a 
Kaluza-Klein type dimensional reduction from 26 to 10 
dimensions. It turns out that the toroidal 
compactification and conversion to fermions is consistent only if 
the lattice that defines the 16-dimensional torus is even and self-dual. 
But it is well-known that there are only two such lattices, called 
$D^+_{16}$ and $E_8+E_8$, and since these have automorphism groups 
$SO(32)$ and $E_8\times E_8$, respectively, one sees at once 
where the origin of these symmetry groups lies. 
The further reduction from 10 to 4 dimensions is, of 
course, another question. One of the more attractive proposals is 
that the quotient, 6-dimensional space, be a Calabi-Yau space~\cite{lo1},
~\cite{lo4}, but 
we do not wish to pursue this question further here. 

\subsection{Gauge Theory and Noncommutative Geometry}

The recent development of non-commutative geometry by Connes~\cite{lo5} has 
permitted the generalization of gauge-theory ideas to the case 
in which the standard  differential manifolds (Minkowskian, 
Euclidean, Riemannian) become mixtures of 
differential and discrete manifolds. The differential operators then 
become mixtures of ordinary differential operators and matrices.
From the point of view of the fundamental physical interactions 
the interest in such a generalization of gauge theory is that the 
Higgs field and its potential, which are normally introduced in an  
ad hoc manner, appear as part of the gauge-field structure. Indeed the 
Higgs field emerges as the component of the gauge potential in 
the 'discrete direction' and the Higgs potential, like the 
self-interaction of the gauge-field, emerges from the square 
of the curvature. The theory also relates to Kaluza-Klein theory because
the Higg's field and potential can 
also be regarded as coming from a dimensional reduction in which the 
discrete direction in the gauge group is reduced to an internal 
direction. 

\subsubsection{Simple Example} 

To explain the idea in its simplest form we follow Connes~\cite{lo5} and use as 
an example the simplest non-trivial case, namely when the continuous 
manifold is a four-dimensional compact Riemannian manifold with 
gauge group $U(1)$ and the discrete manifold consists of just two points. 
With respect to the new discrete (2-point) direction the zero-forms 
(functions) $\omega_0(x)$ are taken to be diagonal $2\times 2$ matrices with 
ordinary scalar function as entries 
\begin{equation}\omega_0(x)=\pmatrix{ f_a(x) &0 \cr 0&f_b(x)}\quad \in 
\quad \Omega_0\,,\label{s813}\end{equation}
$\Omega_0$ denoting the space of zero-forms. The essential new 
feature is the introduction of a discrete component $d$ of the outer 
derivative $d$. This is defined as a self-adjoint off-diagonal 
matrix, i.e. 
\begin{equation}{\bf d}=\pmatrix{0&k\cr k &0}\label{s814}\end{equation}
with constant entries $k$. (More generally one could take the 
off-diagonal elements in $d$ to be complex conjugate bounded 
operators but that will not be necessary for our purpose.) 
The outer derivative of the zero-forms with respect to ${\bf d}$ is obtained 
by commutation 
\begin{equation}{\bf d}\wedge \omega_0\equiv [{\bf d},\omega_0]=(f_b(x)-f_a(x))\pmatrix{0&k\cr 
-k&0}\,. \label{s814b}\end{equation}
The non-commutativity enters in the fact that $d\omega_0$ does not 
commute with the forms in $\Omega_0$. 
The one-forms $\omega_1$ are taken to be off-diagonal matrices 
\begin{equation}\omega_1(x)=\pmatrix{ 0& v_a(x) \cr v_b(x) & 0}\quad 
 \in\quad  \Omega_1\,,\label{s815}\end{equation}
where the $v(x)$'s are ordinary scalar functions and $\Omega_1$ denotes 
the space of one-forms. Note that, according to (\ref{s814b}) the discrete 
component of the outer derivative maps $\Omega_0$ into $\Omega_1$. The outer 
derivative of a one-form with respect to ${\bf d}$ is obtained by 
{\it anti-commutation}. Thus 
\begin{equation}{\bf d}\wedge \omega_1\equiv \{{\bf d},\omega_1\}=(v_b(x)-v_a(x))kI  \in
\Omega_0\,,
\label{s816}\end{equation}
where $I$ is the unit $2\times 2$ matrix. It is easy to check that with 
this definition we have ${\bf d}\wedge {\bf d}\,\wedge =0$ on both $\Omega$-spaces. 

The $U(1)$ gauge group is a zero-form and is the direct sum of the $U(1)$ gauge 
groups on the two sectors of the zero-forms. Thus it has elements of the form 
\begin{equation}U(x)=\pmatrix{ e^{i\alpha(x)} &0\cr 0&e^{i\beta(x)} } \quad \in 
U(1)\,. \label{s817}\end{equation}
Its action on both $\Omega_0$ and $\Omega_1$ is by conjugation. Thus 
under a gauge transformation the zero-forms are invariant and the one-forms 
transform according to 
\begin{equation}\omega_1(x)\rightarrow \omega'_1(x)
=U^{-1}(x)\omega_1(x)U(x)\,.
\label{s818}\end{equation}
Explicitly 
\begin{equation}\omega_1'(x)=\pmatrix{0 & e^{i\lambda(x)} g_a(x)\cr 
e^{-i\lambda(x)} g_b(x) & 0 }\,, 
\quad \hbox{where} \quad
\lambda(x)=\beta(x)-\alpha(x)\,.\label{s819}\end{equation}

The discrete component of a connection takes the form 
\begin{equation}V(x)=\pmatrix{0&v(x)\cr v^*(x)}\label{s819b}\end{equation} 
and thus resembles a hermitian one-form. But, being a connection, it 
is assumed to transform with respect to $U(x)$ as 
\begin{equation}V(x)\rightarrow
V_u(x)=U^{-1}(x)V(x)U(x)+e^{-1}U^{-1}(x)dU(x)\,,\label{s820}\end{equation}
where $e$ is a constant. The transformation law~(\ref{s820}) is the natural 
extension of the conventional transformation law for connection forms.  

The discrete component of the covariant outer derivative is defined to be 
\begin{equation}{\bf D}={\bf d}+eV(x)=\pmatrix{0& k+ev(x)\cr k+ev^*(x)&0}
=e\pmatrix{0& \phi(x) \cr \phi^*(x)&0}\,,\label{s821}\end{equation}
where 
\begin{equation}\phi(x)=v(x)+c\,, \qquad c={k\over e}\,.\label{s822}\end{equation}
The outer derivative with ${\bf D}$ is formed in the same way as with ${\bf
d}$, 
namely by commutation and anti-commutation on the forms $\Omega_0$ 
and $\Omega_1$, respectively. 
From~(\protect\ref{s820}) it follows in the usual way that ${\bf D}$ transforms 
covariantly with respect to the $U(1)$ gauge group, i.e.
\begin{equation}{\bf D}(\phi(x))\rightarrow 
{\bf D}(\phi_\lambda(x))=U^{-1}(x){\bf D}(\phi(x))U(x)\,,\label{s822b}\end{equation}
where
\begin{equation}\phi_\lambda(x)=e^{i\lambda(x)}\phi(x)\,.\label{s823}\end{equation}
This is consistent with the fact that ${\bf D}$ acts on the gauge-group by 
commutation. 

Note that, although the component $v(x)$ of the connection does not 
transform covariantly with respect to $U(1)$, the field $\phi(x)$ does. 
Since $\phi(x)$ is also a space-time scalar, it can therefore be 
identified as a Higgs field. As we shall see, the fact that 
$\phi(x)$ rather than $v(x)$ is identified as the Higgs field is of great 
importance for the Higgs potential.  

Having defined the covariant derivative it is natural to construct 
the curvature. In an obvious notation this can be written as 
\begin{equation}F_{AB}=\pmatrix{F_{\mu\nu} & F_{d\mu}\cr F_{d\mu}
&F_{dd}}\,, 
\label{s824}\end{equation}
where $F_{\mu\nu}$ is the conventional curvature and
\begin{equation}F_{d\mu}=\partial_\mu V-d\wedge A_\mu +[A_\mu,V]=
\pmatrix{0&D_\mu \phi \cr D_\mu\phi^* &0}\equiv D_\mu \Phi\,, 
\label{s825}\end{equation}
where $D_\mu$ is the conventional covariant derivative. 
The interesting component is $F_{dd}$ which turns out to be 
\begin{equation}F_{dd}={\bf d}\wedge V+eV^2\,. \label{s826}\end{equation}
The explicit form of~(\protect\ref{s826}) is 
easily computed to be  
\begin{equation}F_{dd}=\Bigl(k(v+v^*)+evv^*\Bigr)I=e\bigl(\vert\phi\vert^2-c^2\bigr)I\,.
\label{s828b}\end{equation}
Since it is $\phi(x)$ that must be identified as a Higgs field the 
relationship between~(\protect\ref{s828b}) and the standard $U(1)$ 
Higgs potential is obvious. 

Before applying the above formalism to physics, however, we have to introduce 
fermion fields $\Psi(x)$. These are taken to be column vectors of 
ordinary fermions $\psi_a(x)$  
\begin{equation}\Psi(x)=\pmatrix{\psi_a(x)\cr
\psi_b(x)}\,.\label{s829}\end{equation}
The action of the gauge-group and the covariant derivative on them  
is by ordinary multiplication, i.e. 
\begin{equation}U(x)\Psi(x)=\pmatrix{e^{i\alpha(x)}\psi_a(x)\cr 
e^{i\beta(x)}\psi_b(x)}\label{s830}\end{equation}
and 
\begin{equation}{\bf D}\Psi(x)=e\pmatrix{0&\phi(x) \cr \phi^*(x) &0}
\pmatrix{\psi_a(x)\cr\psi_b(x)} 
=e\pmatrix{\phi(x) \psi_b(x) \cr \phi^*(x) \psi_a(x)}\,,
\label{s831}\end{equation} 
respectively. As might be expected from the fact that the fermions 
are $U(1)$-covariant, it is the $U(1)$-covariant field $\phi(x)$, and not 
the component $v(x)$ of the connection, that couples to them in~(\ref{s831}). 

\subsubsection{Application to the Standard Model}

As has already been mentioned the immediate physical interest of the 
non-commutative gauge-theory lies in its application to the standard 
model of the fundamental interactions. The new feature is that it produces 
the Higgs field and its potential as natural consequences of gauge-theory, 
in contrast to ordinary field theory in which they are introduced an ad hoc 
phenomenological manner. The mechanism by which they are produced 
is very like that used in KK reduction so, to put the 
non-commutative mechanism into perspective, let us first 
digress a little to recall the usual KK mechanism.

{\flushleft \bf KK-Mechanism}

Consider the gauge-fermion Lagrangian density 
in $4+n$ dimensions, namely 
\begin{equation}L={1 \over 4}\hbox{tr}\bigl(F_{AB}\bigr)^2 +\bar\psi \gamma^A 
D_A \psi\,,\label{s832}\end{equation}
where $A,B=1....4+n$. If we let $\mu,\nu=0...3$  and $r,s=4...n$ and 
assume that the fields do not depend on the coordinates $x_r$, the 
Dirac operator and the curvature decompose into 
\begin{equation}F_{AB}=\pmatrix{F_{\mu\nu} & D_\mu A_r\cr -D_\mu A_r  & [A_r,A_s] } 
\quad \hbox{and} \quad 
\gamma^A D_A=\gamma^\mu D_\mu+\gamma^r A_r\,,\label{s833}\end{equation}
respectively, and hence the Lagrangian~(\protect\ref{s832}) decomposes into 
\begin{equation}L={1 \over 4}\hbox{tr}\bigl( F_{\mu\nu}\bigr)^2 
-{1 \over 2}(D_\nu A_r)^2+\bar\psi \gamma^\nu 
D_\mu \psi+\bar\psi \gamma^s 
A_s \psi +{1 \over 4}\hbox{tr}\bigl[A_s,A_r]^2\,.\label{s834}\end{equation}
The extra components $A_r$ of the gauge-potential are 
space-time scalars and may therefore be identified as Higgs fields. 
Thus the dimensional reduction produces a standard kinetic term,  
a standard Yukawa term, and a potential for the Higgs fields. The 
problem is that the Higgs potential is not the one required for the 
standard model. In particular, its minimum does not force 
$\vert A_r\vert$ to assume the fixed non-zero value that is necessary 
to produce the masses of the gauge-fields and leptons. 

{\flushleft\bf The Non-Commutative Mechanism}

As we shall now see, the noncommutative mechanism is very 
similar to the KK mechanism. But it eliminates the artificial 
assumption that the fields do not depend on the extra coordinates 
and it produces a Higgs potential that is of the same form as those 
used in standard models. As in the KK case, the procedure is to 
start with the formal gauge-fermion Lagrangian~(\ref{s832}) and expand 
around the conventional 4-dimensional gauge and fermion fields. 

From the discussion of the previous section we see that if we expand 
the Dirac operator and the Yang-Mills curvature in this way we obtain 
\begin{equation}F_{AB}=\pmatrix{F_{\mu\nu} & D_\mu \Phi(x) \cr -D_\mu \Phi(x) & F_{dd} }  
\quad \hbox{and} \quad 
\gamma^A D_A=\gamma^\mu D_\mu+g{\bf D}\,,\label{s835}\end{equation}
respectively, where $g$ is a constant whose value cannot be fixed as 
the theory does not relate the scales of $D_\mu$ and ${\bf D}$. 
The resemblance between~(\protect\ref{s835}) and the corresponding KK expression 
(\protect\ref{s833}) is striking. 

It is clear from (\protect\ref{s835}) that for the non-commutative case the 
formal Yang-Mills-fermion Lagrangian (\protect\ref{s832}) decomposes to 
\begin{equation}L={1 \over 4}\hbox{tr}\bigl(F_{\mu\nu}\bigr)^2 
-{1 \over 2}(D_\nu \phi)^2+\bar\Psi \gamma^\nu 
D_\mu \Psi+G\bar\Psi \Phi \Psi 
+{1 \over 4}\hbox{tr}\bigl(F_{dd}(\phi)\bigr)^2\,, \label{s836}\end{equation}
where 
\begin{equation}\Phi(x)=\pmatrix{0 & \phi(x) \cr  \phi^*(x) &0 }\quad \hbox{and} 
\quad G=eg\,.\label{s837}\end{equation} 
Since the field $\phi(x)$ is a scalar that transforms covariantly with 
respect to the $U(1)$ gauge group it may be interpreted as a 
Higgs field. Hence, in analogy with the KK-mechanism, the 
non-commutative mechanism produces a standard kinetic term, 
a standard Yukawa term and a potential for the Higgs field. The 
difference lies in the form of the potential, which is no longer 
the square of a commutator.  From (\protect\ref{s828b}) we have 
\begin{equation}{1 \over 4}\hbox{tr}\bigl(F_{dd}\Bigr)^2
={1 \over 2}e^2\bigl(\vert
\phi(x)\vert^2-c^2\bigr)^2\,.\label{s837b}\end{equation}
But this is just the renormalizable potential that is used to 
produce the spontaneous breakdown of $U(1)$ invariance. Putting all 
the new contributions together, we see that the 
introduction of the discrete dimension and its associated 
gauge-potential $\phi(x)$ produces exactly the extra terms 
\begin{equation}-{1 \over 2}(D_\mu \phi(x))^2 
+G\bar\Psi(x)\Phi(x)\Psi(x)+{1 \over 2}e^2\bigl(\vert\phi(x)\vert^2-c^2\bigr)^2 
\label{s838}\end{equation}
that describe the Higgs sector of the standard $U(1)$ model. 
Thus, when the concept of manifold is generalized in the manner 
dictated by non-commutative geometry, the standard Higgs sector 
emerges in a natural way. Note, however, that since there 
are three undetermined parameters in~(\ref{s836}), the non-commutative 
approach does not achieve any new unification in the sense of  
reducing the number of parameters.
However, it considerably reduces the ranges of the parameters, places strong
restrictions on the matter-field representations and even rules out the
exceptional groups as gauge groups~\cite{63n}. 

Of course the above model is only a toy one since it uses the 
gauge-group $U(1)\times U(1)$ rather than the gauge-groups $U(2)$ and  
$S(U(2)\times U(3))$ of the standard electroweak and electroweak-strong 
models or the gauge-groups of grand unified theory. 

However, the general 
structure provided by non-commutative geometry can be applied to any 
gauge group. Connes himself~\cite{lo5} has applied it to the standard model.
There is some difficulty in applying it to grand unified theories
because of the restrictions on fermion representations, but a modified
version has been applied to grand unified theories in~\cite{lo6}.
As in the toy model, the non-commutative approach does not achieve any new 
unification in the sense of reducing the number of parameters, though,
as already mentioned, it introduces some important restrictions. Most
importantly, it provides a new and interesting interpretation.

\begin{acknowledgments}
We are indepted to C.N.~Yang for important remarks which improved the paper.
A number of people gave us positive reactions and welcome comments. We
thank, in particular, D.~Giulini, F.~Hehl, U.~Lindstrom, T.~Schucker and
D.~Vassilevich.
\end{acknowledgments}

\end{document}